\documentclass[12pt]{article}
\usepackage[a4paper, total={6.6in, 9.1in}]{geometry}

\usepackage[natbib=true,
            style=numeric-comp,
            citestyle=numeric-comp, 
            backend=biber,
            maxnames=999, 
            uniquelist=false,
            uniquename=false,
            date=year,
            maxcitenames=2,
            sorting=none
        ]{biblatex}
\usepackage{graphicx}
\usepackage{subcaption}
\usepackage{xspace}
\usepackage[dvipsnames]{xcolor} 
\usepackage{siunitx}
\usepackage{amssymb,amsthm}
\usepackage{empheq}
\usepackage{xcolor}
\usepackage{xargs}
\definecolor{linkcolor}{rgb}{0,0,0.5} 

\usepackage[colorlinks=true,
    urlcolor=linkcolor, 
	linkcolor= linkcolor, 
  citecolor=linkcolor, 
  linktocpage=true,
]{hyperref}
\usepackage{cleveref}

\graphicspath{ 
{FiguresHL/}{standalone/}}

\newtheorem{rmk}{Remark}

\newcommand{\tI}[1]{\mathbf{#1}}

\newcommand{\tII}[1]{\mathcal{\mathbb{#1}}	}

\newcommand{\ie}{\emph{i.e.}\xspace}
\newcommand{\eg}{\emph{e.g.}\xspace}
\newcommand{\cf}{\emph{c.f.}\xspace}
\newcommand{\kos}{\texorpdfstring{$k-\omega$}{k-w} SST\xspace}

\newcommand{\id}{\tII{I}}
\newcommand{\divs}[1]{\nabla\cdot #1}
\newcommand{\vv}{\tI{u}}
\newcommand{\grads}[1]{\mathbf{\nabla} #1}
\newcommand{\rots}[1]{\nabla \times #1}
\newcommand{\crossop}{\otimes}
\newcommandx{\partialII}[3][1=]{\dfrac{\partial^{#1} #2}{\partial #3^{#1} } }
\newcommandx{\dt}[2][1=]{\partialII[#1]{#2}{t}}
\newcommandx{\dn}[2][1=]{\partialII[#1]{#2}{n}}

\newcommand{\widebar}[1]{
    \mskip.5
        \thinmuskip
            \overline{
                \mskip-.5
                \thinmuskip {
                    #1
                    } 
                \mskip-.5
                \thinmuskip
    }
    \mskip.5
    \thinmuskip
} 
\newcommand{\mean}[1]{ \widebar{ #1\rule{0pt}{2.25mm} } }
\newcommand{\norm}[1]{\left\lVert#1\right\rVert} 

\newcommand{\openfoam}{OpenFOAM\textregistered{}\xspace}

\newcommandx{\sfig}[3][1=,3=0.49]{
        \begin{subfigure}[c]{#3\textwidth}
            \centering
            \includegraphics{#2}
                \caption{#1}
                \label{fig:sub:#2}
        \end{subfigure}%
}

\usepackage{listings}
\def\ofkw{\lstinline[
  breaklines=true ,
  breakatwhitespace=false,
    morekeywords={},
  basicstyle={\small\ttfamily},
  numbers=left,
  numberstyle=\tiny\color{gray},
  keywordstyle=\color{blue},
  commentstyle=\color{dkgreen},
  stringstyle=\color{purple},
]}

\providecommand{\keyword}[1]
{
  \small	
  \textbf{\textit{Keywords---}} #1
}

\DeclareSIUnit{\kunit}{\square\meter\per\square\second}
\newcommand{\ti}[2]{\SI{#1}{\hour}~\SI{#2}{\minute}} 

\usepackage{authblk}

\usepackage{multirow}

\title{Assessment of one-way coupling methods from a potential to a viscous flow solver based on domain- and functional-decomposition for fixed submerged bodies in nonlinear waves
  \footnote{ published as a research article in European Journal of Mechanics - B/Fluids,
  Volume 95, September–October 2022, Pages 315-334. doi:  \href{https://doi.org/10.1016/j.euromechflu.2022.05.011}{https://doi.org/10.1016/j.euromechflu.2022.05.011}}
}
\author[1,2]{Fabien Robaux} 
\author[1,2,3]{Michel Benoit} 

\affil[1]{ Aix Marseille Univ, CNRS, Centrale Marseille, Institut de Recherche sur les Ph\'enom\`enes Hors-Equilibre (IRPHE), UMR 7342, Marseille, France}
\affil[2]{Saint-Venant Hydraulics Laboratory (Ecole des Ponts ParisTech, EDF R\&D), Chatou, France}
\affil[3]{EDF R\&D, Laboratoire National d’Hydraulique et Environnement (LNHE), Chatou, France}

\date{}

\begin{document}

\maketitle

\vspace{-1.2cm}
\begin{abstract}
  To simulate the interaction of ocean waves with marine structures, coupling approaches between a potential flow model and a viscous model are investigated.
  The first model is a fully nonlinear potential flow (FNPF) model based on the Harmonic Polynomial Cell (HPC) method, which is highly accurate and best suited for representing long distance wave propagation. The second model is a CFD code, solving the Reynolds-Averaged Navier-Stokes (RANS) equations within the \openfoam toolkit, more suited to represent viscous and turbulent effects at local scale in the body vicinity.
Two one-way coupling strategies are developed and compared in two dimensions, considering fully submerged and fixed structures. 
A domain decomposition (DD) strategy is first considered, introducing a refined mesh in the body vicinity on which the RANS equations are solved. Boundary conditions and interpolation operators from the FNPF results are developed in order to enforce values at its outer boundary.
The second coupling strategy considers a decomposition of variables (functional decomposition, FD) on the local grid. As the FNPF simulation provides fields of variables satisfying the irrotational Euler equations, complementary velocity and pressure components are introduced as the difference between the total flow variables and the potential ones. Those complementary variables are solutions of modified RANS equations.
Extensive comparisons are presented for nonlinear waves interacting with a horizontal cylinder of rectangular cross-section. The loads exerted on the body computed from the four simulation methods (standalone FNPF, standalone CFD, DD and FD coupling schemes) are compared with experimental data. It is shown that both coupling approaches produce an accurate representation of the loads and associated hydrodynamic coefficients (inertia and drag) over a large range of incident wave steepness and Keulegan-Carpenter number, for a small fraction of the computational time needed by the complete CFD simulation. 
\end{abstract}

\keyword{
Fluid-structure interaction, 
Coupling methods, 
Potential-viscous coupling, 
Domain decomposition, 
Functional decomposition, 
Nonlinear water waves
}


\section{Introduction} \label{sec:intro}
To simulate the propagation of ocean waves and wave interaction with marine or coastal structures various modeling approaches can be invoked, depending on the assumptions that can be made. In the most general case, the surface waves have to be considered as nonlinear and dispersive, the flow possesses a rotational part with viscous and turbulent effects present, at least in the vicinity of bodies and domain boundaries (\eg seabed). Furthermore, in case of wave breaking or in the process of wave interaction with a floating structure, air entrainment can occur, so that mixed air-water flow has to be considered, with varying properties of the fluid characteristics (\eg density, viscosity). An accurate modeling of all these features at the spatial scale of kilometers in practical applications remains however challenging. An extensive literature review on modeling wave-structure interaction (WSI) was recently given by \textcite{davidson_efficient_2020} where many existing approaches are discussed from, but not limited to, the wave energy converter modeling point of view. 

Computational Fluid Dynamics (CFD) approaches, aiming at solving the Navier-Stokes (NS) equations -- or more frequently the Reynolds Averaged Navier-Stokes (RANS) equations -- for two-phase flows are able to capture most of the above mentioned effects and have made impressive progress in the recent years, both in terms of accuracy and efficiency \parencite[see, \eg][]{jacobsenFuhrmanFredsoe2012,kim2016numerical,oggiano2017reproduction}. 
Those CFD approaches are particularly well suited for simulating WSI processes, in particular viscous, turbulent and rotational effects, and possible air entrainment. Note that, in the context of WSI, these effects are mostly contained in the close vicinity of the body.

At a larger scale, when considering wave propagation over a domain covering several dozens of wavelengths, it is often possible to neglect viscous effects and to consider the flow as irrotational, which permits to use a potential flow approach. Many models have been developed for decades under this assumption, leading to the so-called Fully Nonlinear Potential Flow (FNPF) theory if full nonlinearity of the free surface boundary conditions (BC) is kept \parencite[\eg][]{tavassoli2001interactions}, partially or weakly nonlinear models \parencite{pinkster1980low,philippe:hal-01198807}, and simplified linear potential flow models \parencite[\eg][]{lee1995wamit,ansys2013aqwa,babarit2015theoretical}. These models, in particular the FNPF ones, have demonstrated both higher accuracy and lower computational resource requirements in comparison with CFD codes to simulate the large-scale propagation phase, because of their intrinsic lower numerical dissipation rate. 

As a consequence, an intuitive idea is to apply each model over a domain and at a scale where it best performs, namely employing the CFD approach where needed in the vicinity of the body, and the FNPF approach far from the body to benefit from a more accurate and cheaper approach to simulate long distance wave propagation. This type of methods is referred to as a ``coupled'' or ``hybrid'' method.
The idea of using different models to best capture different physical effects is not new: seminal studies that tried to include the boundary layer viscous effects, within an otherwise inviscid simulation, can be seen as a hybrid model. For example, the very first boundary layer theory \parencite{prandtl_uber_1904} falls into that category, as well as the further work by \textcite{lighthill_displacement_1958}.
The introduction in the context of wave flows was done by \textcite{dommermuth_numerical_1997}, who applied a decomposition of the flow into irrotational and rotational parts to solve the contact line problem in bow waves.

Over the last two decades, several authors have successfully developed coupling schemes that use each model in the area where it is most adequate. Those coupling schemes can be separated into two main categories: Domain Decomposition (DD) methods and Functional Decomposition (FD) methods. The first one (DD) uses two different mathematical models and solution methods applied on distinct domains. In most cases, the domains do not overlap, and information is exchanged between the models only at common boundaries. A variant of this approach is to introduce overlapping zones at the interface between the two domains, with a progressive matching of BCs over the extent of this zone. A comprehensive review on DD model coupling in the context of WSI was recently given by \textcite{di_paolo_wave_2021}. This approach is further discussed in \cref{sec:domainCoupling:literature}.
The second type of decomposition (FD) leads to a modification of the equations themselves. A flow that is solution of a simplified set of equations, denoted ``model A'', already verifies a significant portion of the more generic equations, denoted ``model B'' (in our case, models A and B will be Euler and NS equations respectively). Thus, if model A is solved first over the whole domain of interest, solving model B from scratch can be seen as largely redundant. Modifying the equations of model B to take into account the already computed part by model A results in a modified model B$^*$ whose solution is hoped to be easier and  significantly cheaper in terms of computational cost.
In this framework, the domain where model B$^*$ is solved needs to be a sub-domain of the domain where model A is solved, and we state that $\vv_B = \vv_A + \vv_{B^*}$, where $\vv_B$ is the total velocity solution of model B, and $\vv_A$, $\vv_{B^*}$ the velocities computed by the model A and B$^*$ respectively (a similar decomposition is applied to other variables, \eg pressure). This approach is further discussed in \cref{sec:velocityCoupling:literature}.

Another distinction can be made between one- and two-way coupling methods (also referred to as weak and strong couplings). Within a one-way framework, the first model is used to solve a given wave field. The second model imports the (boundary or volume) values to solve in a more contained zone the more complex set of equations. No feedback from the second model to the first one happens, which implies that the first one is independent and can be run \emph{a priori}. In a two-way coupling method, both models receive information from the other (at their common BCs on a DD approach), each one having an effect on the other. Thus, they have to evolve simultaneously and ``wait'' for each other over the duration of the simulation.

In the present study, both DD and FD approaches are tackled and compared to simulate WSI for ocean engineering applications. However, as a first step, only one-way coupling is considered. Furthermore, the coupling methodologies are evaluated here with a fixed submerged body. The remainder of this article is organized as follows: in \cref{sec:review}, a literature review on DD and FD methods is proposed. The potential and viscous models used are briefly presented in \cref{sec:models}. The DD and FD coupling methods developed and implemented during this work are presented in \cref{sec:domainCoupling} and \cref{sec:velocityDecomposition} respectively. In \cref{sec:resultsModelsComp} we report and discuss a series of tests and sensitivity studies to assess and compare the requirements and performances of the two coupling methods. Then, in \cref{sec:resultsVsLiterature}, we present the results of four modeling approaches (namely a FNPF one, a fully viscous RANS one, and the DD and FD coupling schemes) applied to simulate the interaction of waves with a submerged horizontal cylinder of rectangular cross-section. The results are compared with experimental data from several sources covering a range of incident wave conditions. Finally, conclusions are summarized in \cref{sec:conclusion}.

\section{Literature review on coupling methods} \label{sec:review}
A  bibliographic review of DD and FD coupling approaches is given in the following sub-sections. As these approaches have been extensively developed over the last 25 years, only some of the notable works are listed below. This list of references is not exhaustive due to the variety of possible approaches and variants.

\subsection{Overview of DD coupling methods} \label{sec:domainCoupling:literature}
DD methods were first applied in the aerodynamics field \parencite{lock_viscous-inviscid_1987}, and then introduced in the context of ship sea-keeping studies more than two decades ago by \textcite{campana1994domain,campana1995viscous}. 
An early attempt to use a DD method in the context of moored ship was presented in \textcite{bingham_hybrid_2000}. \textcite{quemere_new_2001} suggested a coupling approach for highly different block discretizations and applied it to a RANS/LES (Large Eddy Simulation) coupling in \textcite{quemere_zonal_2002}.

As summarized in the review of \textcite{di_paolo_wave_2021} (see in particular their Table 1 which offers a synthetic view of a large set of approaches), many types of models can be interfaced. For example, wave theory based models can be used in conjunction with potential or RANS models, see \eg \textcite{wei_cost-effective_2017} or \textcite{christensen_transfer_2009} for applications of DD to the coupling of a Boussinesq wave model and a RANS solver, while \textcite{kassiotis_coupling_2011} used this approach to couple a Boussinesq wave model with a Smoothed Particle Hydrodynamics (SPH) solver.

Another widely used combination is the matching of a potential flow solver with NS or RANS equations. The potential equations can for example be solved with a High-Order Spectral (HOS) approach \parencite{choi_generation_2018} or with a Boundary Element Method (BEM) \parencite{colicchio_bem-level_2006}. Recently, \textcite{HPC:hanssen2019non,siddiqui_validation_2018} coupled NS equations with a potential solver based on the Harmonic Polynomial Cell (HPC) method.
\textcite{siddiqui_validation_2018} studied the wave interaction with a damaged ship and the resulting hydrodynamics, for which they later released two experimental studies \parencite{siddiqui_experimental_2019,siddiqui_experimental_2020}.

In \textcite{sriram_hybrid_2014}, the FNPF model, solved with a BEM, is strongly coupled with a NS solver based on a Finite Element Method (FEM), and applied to a wave breaking problem. This numerical model is later applied in \textcite{kumar_hybrid_2020} to the estimation of long wave run-up.

A solver denoted ``qaleFOAM'' was also developed, coupling a FNPF solver using a Quasi Lagrangian Eulerian FEM with \openfoam. The method is presented and applied in \textcite{li_zonal_2018,li_numerical_2018,yan_numerical_2019,wang_numerical_2020}. 

Another innovative DD method, presented in \textcite{kristiansen_gap_2012,kristiansen_validation_2013}, employs a linear potential model as the external model, that is responsible of solving for the free surface evolution. The coupling is then made with a NS based model and both are solved with the Finite Volume Method (FVM).

Once again, more literature references can be found in \textcite[section 4]{davidson_efficient_2020} and \textcite{di_paolo_wave_2021}.

\subsection{Overview of FD coupling methods} \label{sec:velocityCoupling:literature}
The acoustic and aerodynamic field of research was the first to consider such type of variable or field decomposition \parencite{morino_helmholtz_1986,morino_toward_1994,morino_new_1999,hafez2006numerical,hafez2007improved}, mostly applied to solving the boundary layer problem where the viscosity plays a major role.

An actively studied and applied methodology in the last two decades was proposed by \textcite{SWENSE:ferrant2003potential,SWENSE:gentaz2004numerical,SWENSE:luquet2004viscous}. The velocity, pressure and surface elevation are decomposed into their incident and diffracted/radiated components. A potential theory based wave model is used to explicitly obtain the incident field (ignoring the presence of bodies in the domain), and then a modified version of the RANS equations, denoted the Spectral Wave Explicit Navier-Stokes Equations (SWENSE), is derived. Those equations require the explicit values of the incident fields (such as the potential wave elevation, potential velocities): given a grid at a particular time instant, the potential fields and kinematics are calculated. Afterwards, the newly derived SWENSE equations are solved to yield the diffracted component. 
Notice that the potential-viscous coupling scheme is one-way in the sense that no feedback effect on the potential model is at play. This, however, does not come with any hypothesis but instead with the drawback of having to mesh the fluid domain up to relatively far from the body of interest: %
the diffracted fields do not vanish when the distance to the object increases (in 2D inviscid cases). 
Nevertheless, the main advantage of the method is to reduce the complexity of the BCs for the SWENSE equations, as no incident wave theory has to be imposed at the SWENSE BC. Thus, only a damping of the diffracted/radiated wave field has to be set. The problem often encountered with RANS simulations addressing the propagation of incoming waves without significant damping is also avoided. 
\textcite{luquet_applications_2007} described the methodology and the numerical implementation in great detail. They also introduced an irregular incident wave potential simulation, belonging to the family of HOS schemes. 
Over the years, numerous test cases were investigated with successful results. 
For example, \textcite{luquest2007simulation} focused on  a tension leg platform, \textcite{alessandrini_numerical_2008,monroy_rans_2009} applied the method for ship sea-keeping, and \textcite{li_calculation_2017} solved the SWENSE equations on a vertical cylinder piercing the free surface. 
A development of the SWENSE equations within the OpenFOAM package employing a Volume of Fluid (VoF) method and thus a two-phase flow was recently done by \textcite{vukcevic_decomposition_2016-1,vukcevic_decomposition_2016,vukcevic_numerical_2016}. In a similar manner, while this method was originally developed within a single phase RANS solver, recent developments were conducted on the use of a fixed grid solver first with the use of a level-set tracking method \parencite{reliquet_simulation_2013,reliquet_simulations_2019} and more recently with the use of the VoF method within a two-phase solver \parencite{li_challenges_2018,li_progress_2018,li_spectral_2021}. 

Another approach was developed by \textcite{Beck:kim2005complementary} in which the potential model is solved with the presence of the body. The complementary velocity is defined as $\vv^*=\vv_t - \vv_p$, where $\vv_t$ is the total velocity field satisfying the NS equations and $\vv_p$ is the potential one satisfying the Laplace equation. The same decomposition is performed on the pressure field.
Turbulent and rotational effects are thus comprised into these newly defined fields.
In the potential solver developed by \textcite{Beck:kim2005complementary}, a Rankine source type method is used. The studied cases mostly lie in the aerodynamic field of research. 

With another approach, \textcite{edmund_improved_2011,Beck:edmund2012velocity,Beck:edmund2013velocity} also decompose the velocity field into a potential and a rotational components. With the objective of allowing for a truncation of the NS domain, the potential part is sought as a solution of the Laplace equation (without invoking the non-viscous assumption). 
The RANS sub-problem does not differ from the original one except at its outer boundary on which the potential velocity is imposed. %
Reciprocally, the potential model solves the Laplace equation with a special coupled boundary on the body surface leading to a retroactive effect of the NS sub-problem onto the potential one. 
An iterative method is set to achieve convergence and consistence of the two sub-problems on the coupling boundaries. An important domain reduction is achieved on a steady flow over a NACA profile airfoil. This method was extended to steady free surface flows in \textcite{rosemurgy_velocity_2012}. Recently, work has been conducted to develop the unsteady version of this approach by \textcite{chen_velocity_2015}, further extended to 3D and applied to the Wigley hull by \textcite{chen_velocity_2017}.

\textcite{grilli_modeling_2009,harris_coupling_2010,harris_perturbation_2012} developed a 2D one-way coupling scheme with a LES model in order to study the wave-induced transport of sediments near the sea bottom. A FNPF wave tank is used to compute the overall wave field while perturbed NS-LES equations are solved in order to capture the fine scale viscous effects in the boundary layer zone. Later, a coupling between a FNPF model (solved with a BEM) and a NS Lattice-Boltzmann was developed basing on the same perturbation approach \parencite{janssen_modeling_2010,oreilly_hybrid_2015}.

In the context of vortex induced vibrations, \textcite{li_hybrid_2017} recently adopted a FD approach to couple a simplified NS solver (denoted Quasi-turbulent model) with another more complicated RANS model. In the present work, modifications are made on the equations of the second model, in a similar manner as \textcite{Beck:kim2005complementary}. Note that in their method, the two models are based on the NS equations, while only the second one solves for the turbulent variables. The turbulent viscosity can thus affect the first model and its solution. 

In a similar manner as \textcite{Beck:kim2005complementary}, \textcite{zhang_multi-model_2018} decomposed the total fields into a potential part, solved with a two-phase Euler solver, and complementary fields for which complementary RANS equations are derived. The coupling is done in a one-way manner but achieves a domain size reduction and improved stability by making use of transition zones (also called relaxation zones) to match the solution at the interface of the two domains.

\section{Presentation of the CFD and potential models} 
\label{sec:models} 
In this section we briefly present the RANS CFD model (\cref{sec:CFD}) and the FNPF model (\cref{sec:FNPF}) that will be used either as standalone tools to model WSI, or in a coupled mode using the DD and FD coupling methods presented in \cref{sec:domainCoupling} and \cref{sec:velocityDecomposition} respectively.
\subsection{The RANS CFD model} \label{sec:CFD}
The CFD model employed in this study is based on the RANS-VoF solvers available in the \openfoam toolbox. The associated mathematical models and numerical techniques are briefly discussed hereafter. Further details can be found in \eg \textcite{jasak_error_1996}.

\subsubsection{RANS equations} \label{sec:RANS}
For an incompressible Newtonian fluid, considered hereafter, the Cauchy equations can be expressed under the form of the classical Navier-Stokes equations:
\begin{subequations}
    \begin{empheq}[left=\empheqlbrace]{align}
        & \divs \rho \vv = 0\label{eq:NSmaindivT} \\
        & \dt{\rho \vv} + \divs(\rho \vv\crossop\vv) = - \grads p  + \divs \tII{T} + \tI{f} \label{eq:NSmainMomT}
    \end{empheq}
    \label{eq:NSmainSystemT}
\end{subequations}
where $\tI{u}=(u_x, u_y, u_z)^T$ is the velocity vector, $\rho$ the fluid density -- which can vary in a two-phase flow --, $p$ is the pressure, $\crossop$ is the outer product and $\tII{T}$ the shear stress tensor: $\tII{T} =  \mu \left[ \grads \vv + \left(\grads \vv\right)^T \right] - \dfrac{2}{3} \mu \divs \vv \ \id $,  with $\id$ the identity matrix and $\mu$ the dynamic viscosity of the fluid.

In order to reduce the computational cost, the RANS method is employed. First, the equations are statistically averaged, by splitting the velocity into its mean and fluctuating components $\vv = \mean{\vv} + \vv'$.  The Reynolds stress tensor ($\rho \mean{\vv'\crossop \vv'} $) is modeled by assuming a linearity of the latter with respect to the shear stress tensor of the mean flow field, using the so-called Boussinesq assumption:
\begin{equation}
    \begin{aligned}
        -\rho \mean{\vv'\crossop\vv'} & = \dfrac{\mu_t}{\mu} \mean{\tII{T}} -\dfrac{2}{3}\rho k \id\\
                          & = 
    \mu_t \left[ \grads{\mean{\vv}} + \left(\grads\mean{\vv}\right)^T \right]
    -\dfrac{2}{3}\mu_t \divs \mean{\vv} \ \id
    -\dfrac{2}{3}\rho k \id
    \end{aligned}
    \label{eq:boussinesqAssumption}
\end{equation}
where $k=\dfrac{1}{2} \mean{\vv' \cdot \vv'}$ is the turbulent kinetic energy (TKE), and $\mu_t$ the turbulent viscosity. 

Eventually, the RANS equations, governing the dynamics of mean flow variables, are given by:
\begin{subequations}
    \begin{empheq}[left=\hspace{-0.3cm}\empheqlbrace]{align}
    & \divs \rho \mean \vv = 0  \label{eq:RANSequations:div}\\
    & \dt{\rho \mean{\vv} } + \divs{ \rho\mean{\vv}\crossop\mean{\vv} } = - \grads{\mean{p}}         
     -\divs{
             \underbrace{
            \left(
            \mu_{\text{eff}}\left[ 
                \grads{\mean{\vv}}   
                +\left(\grads\mean{\vv}\right)^T 
            \right]
            -\dfrac{2}{3} \mu_{\text{eff}} \divs \mean{\vv} \id
            -\dfrac{2}{3}\rho k \id 
            \right)}
            _{\tII{T} _{\text{eff}}(\mean{\vv})}
        }
        +\tI{\mean{f}} \label{eq:RANSequations:Mom}
    \end{empheq}
\end{subequations}
where the effective viscosity $\mu_{\text{eff}}$ is defined as $\mu_{\text{eff}}=\mu+\mu_t$.
In that framework, two new variables have been incorporated into the problem ($k$ and $\mu_t$). Thus, closure equations are required for this system, they are referred to as turbulence models. For example,  $k-\epsilon$ or $k-\omega$ models are two-equation closure models while the Spalart-Allmaras model \parencite{spalart_one-equation_1992} is a one-equation closure model. In this study, the $k-\omega$ SST model of \textcite{menter_two-equation_1994} will be used, more specifically the modified version provided by \textcite{devolder_application_2017}. 
Note that in the following, the overbar symbol on mean flow variables will be omitted for simplicity, and because only the mean velocity and mean pressure will be used in the CFD context.

\subsubsection{Volume of Fluid (VoF) method for free surface tracking} \label{sec:VoF} 
In order to capture and simulate the evolution of the interface between the two non-miscible fluids, the VoF approach is selected \parencite{hirt1981volume}. 
With this method, the equations are represented in a continuous way across the interface:
a scalar function that represents the volume fraction of a given fluid in a given computational cell is defined. Thus, the two fluids are represented and solved as one with varying physical properties. In \openfoam, the volume fraction, denoted $\alpha$, is a function of $t$ and $\tI{x}$. Thus, when this scalar equals unity in a given cell, this cell is full of fluid 1, \ie water in our case. If it is zero, this cell is full of fluid 2, \ie air in our case. Any value of $\alpha$ in between means that the free surface crosses this cell and gives the relative fraction of water inside this cell.
Transport equation of this volume fraction is, taking into account the incompressibility of the fluids, given by:
\begin{equation}
  \dt{\alpha} + \grads\cdot{(\alpha  \vv) } + \grads\cdot ( \alpha (1-\alpha)\vv^r ) = 0
    \label{eq:advectionAlpha}
\end{equation}
where $\vv^r$ is a numerical relative velocity whose main role is to compress the interface. Further details can be found, for example, in \citet{deshpande_evaluating_2012}.

\subsubsection{Finite Volume Method (FVM)} \label{sec:FVM}
The discretization of the equations in both time and space are done employing the FVM. Given a quantity $q$, in every control volume $V_P$ which centroid is denoted $P$ and at every time $t$, and time step $\Delta t$, the ``semi-discretized'' equation should be satisfied:
\begin{equation}
    \int_t^{t+\Delta t}{}
    \left[
            \underbrace{ \left.  \dt{\rho q}\right|_P V_P }_\text{Time derivative}
            +\underbrace{\sum_{faces}F_f q_f }_\text{Advection}
            +\underbrace{\sum_{faces} \tI{S} \rho_f \Gamma_f (\grads q)_f }_\text{Diffusion}
        \right] dt
    =
    \int_t^{t+\Delta t}{}
    \underbrace{(S_0V_p+S_1V_p q_p)}_\text{Source}
        dt
    \label{eq:semiDiscretizedQ}
\end{equation}
where the source term $S$ has been linearized, $F_f$ is the face flux, $\tI{S}$ the face normal vector, and $\Gamma$ the diffusion parameter. Quantities are indexed with $\cdot_f$ when evaluated at a face and $\cdot_P$ when evaluated at the cell centroid.

\subsubsection{The coupled velocity-pressure system} \label{sec:coupledVelocityPressure}
Applying the discretization presented in \cref{eq:semiDiscretizedQ} to the RANS momentum \cref{eq:RANSequations:Mom}, assuming that the volume force reduces to only the gravitational force, yields the following form:
\begin{equation}
    \vv = \dfrac{1}{\tII{A}} ( \tII{H}(\vv) + \tI{R} 
    \underbrace{- \grads p + \rho \tI{g}}
_{-\grads p_d} )
    \label{eq:HandAonU}
\end{equation}
where $\tII{A}$ and $\tII{H}$ are respectively the diagonal and off-diagonal part of the obtained matrix,  $\tI{R}$ is the source term, containing all explicit contributions (\eg values at the beginning of the time-step), and $p_d = p -\rho g z$ denotes the dynamic pressure. From \cref{eq:HandAonU}, one could derive the Poisson equation for the dynamic pressure, under the form:
\begin{equation}
    \divs{\left(\dfrac{1}{\tII{A}} \grads p_d  \right)} = \divs { \dfrac{\tII{H}(\vv) +\tI{R} } {\tII{A}}} 
    \label{eq:pEqn}
\end{equation}

Together, \cref{eq:HandAonU,eq:pEqn} form the velocity-pressure coupled system that is to be solved. From \cref{eq:pEqn}, $p_d$ is computed from a given velocity field $\vv$, then from the newly computed pressure, a new velocity field is computed explicitly from \cref{eq:HandAonU}. Repeating this procedure until convergence allows to obtain a solution $(\vv, p_d)$ of the coupled system. Because $\tII{H}(\vv)$  is nonlinear with respect to $\vv$, one could update the coefficients of the off-diagonal part of the matrix whenever $\vv$ is updated. This is classically not done, thus neglecting the nonlinear coupling before the velocity-pressure coupling. However, encapsulation algorithms such as for example the \ofkw{PIMPLE} method can be used to counteract this shortcoming.

\subsection{The fully nonlinear potential flow (FNPF) model} \label{sec:FNPF}
The FNPF  model used in this study to generate and propagate the water waves, as well as computing the diffraction/reflection effects due to the presence of a structure, is based on the Harmonic Polynomial Cell (HPC) method introduced by \textcite{HPC:shao2012towards,HPC:shao2014fully,HPC:shao2014harmonic}. The current model was developed and implemented in two dimensions (2D) $(x,z)$ by \textcite{robaux2018modeling,robaux_numerical_2020,robaux_development_2021-1}.  

In this approach, a numerical cell is defined by assembling in 2D framework four adjacent quadrilateral cells in a quadrangular mesh. A cell comprises one center node (denoted with local number 9) and 8 exterior nodes lying on its boundary (with local numbers from 1 to 8). The potential is then approximated in each cell as a weighted sum of the potentials at those 8 exterior nodes: 
  \begin{equation}
    \phi(\tI x)= \sum_{i=1}^{8}  \left(\sum_{j=1}^{8} C^{-1}_{ji}  f_j(\tI{\bar{x}}) \right) \phi_i \label{eq:mainwithcij}
  \end{equation}
    where $C_{ij}$ are the coefficients of a $8\times8$ matrix, only dependent on the geometry and $f_j(\tI{\mean{x}})$ are the eight first harmonic polynomials ($f_1(x,z) =1$, $f_2(x,z) =x$, $f_3(x,z) =z$, $f_4(x,z) =xz$, \emph{etc}.) applied at the location of interest, relative to the cell center $\tI x_9$ (\ie $\tI{\bar{x}} = (x,z) =\tI x - \tI x_9$). As each $f_j$ polynomial is a fundamental solution of the Laplace equation, so is the approximation \cref{eq:mainwithcij}. For every cell, \cref{eq:mainwithcij} is applied at its center $\tI x_9$, yielding an equation for every node located in the bulk of the water domain, which participates to the global linear system of unknowns $\phi_i$. This system is then closed by invoking the different BCs of Neumann and Dirichlet type \parencite{HPC:shao2014harmonic,robaux_development_2021-1}. It is then solved with a preconditioned GMRES  iterative solver from \textcite{gmres:saad2003iterative} modified as suggested by \textcite{baker_simple_2009} to yield the values of the potential at all nodes of the grid.

In our approach, all FNPF-HPC runs are performed taking into account the presence of the body by defining a mesh fitted to the body boundaries on which we also resolve the Laplace problem. Communications between the ``background'' mesh and the ``body fitted'' mesh are done \emph{via} implicit interpolations based on \cref{eq:mainwithcij}. In the ``body fitted'' mesh, Neumann conditions are enforced at the nodes lying on the physical body boundaries. This ensures that the impermeability condition is fulfilled.%

The corresponding flow velocities and pressure are thus available in the whole actual fluid domain, even though they neglect both viscous and turbulent effects. The present implementation of the HPC method yields accurate pressure fields, using the Bernoulli equation. The accuracy is obtained thanks to the formulation and solution of a second Laplace problem on the time derivative of the velocity potential $\phi_{,t}=\dt{\phi}$. 
For further details regarding the method, the reader is referred to \textcite{robaux_development_2021-1, robaux_numerical_2020}.

Potential loads can thus be obtained by integrating the pressure along the body wetted boundaries from the FNPF-HPC simulation prior to running the RANS solver. Results from the FNPF-HPC method will be discussed and compared to the coupling approaches results in \cref{sec:resultsModelsComp,sec:resultsVsLiterature}.

In this work, only one-way coupling is studied. Thus, all computations with the FNPF-HPC model are done \emph{a priori}, and the current implementation of the coupling algorithms builds a solution on top of the HPC solution without influencing it. 

Although we only present hereafter results obtained using this FNPF-HPC model as external solver, the present coupling schemes have been implemented so as to be able to use any external results available in an \openfoam formatted case.

\section{The Domain Decomposition (DD) approach}
\label{sec:domainCoupling}
\begin{figure}[htbp!]
  \begin{center}
    \includegraphics{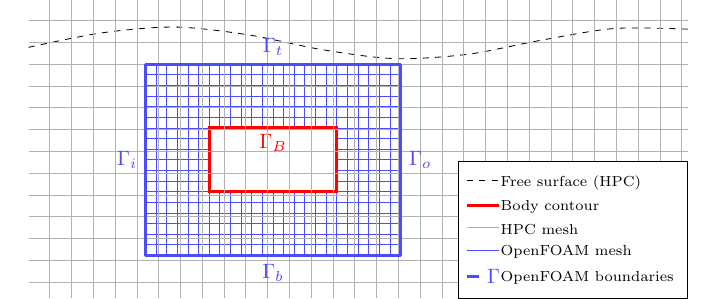}

    \caption{Schematic representation of the DD coupling method applied here. Note that only one mesh is represented for each model here. In our application, two meshes will be used for the potential solution, corresponding to the different HPC grids, namely ``background'' and ``body fitted'' such that impermeability is enforced at the body boundary in the potential model.}
  \label{sch:coupling}
  \end{center}
\end{figure}

The DD method consists in splitting the spatial domain into two separate regions and attributing to each region a given mathematical model (see \cref{sch:coupling}). For the problem of interest, as discussed in \cref{sec:intro}, it is well known that FNPF are well suited to simulate wave propagation and the FNPF-HPC model presented in \cref{sec:FNPF} is used for that purpose. On the other hand, as this model neglects viscous and rotational effects, turbulent effects are not captured and the physics of the flow in the vicinity of a body is oversimplified. Thus, in a local area around the body, a turbulent RANS VoF model presented in \cref{sec:CFD} is set up and developed within the \openfoam framework. We restrict here the study to a fully submerged body, such that no coupling in terms of volume fraction is needed.

\subsection{Spatial and temporal interpolations}
In the DD approach, the equations within the bulk of the local (viscous) domain are not modified. Thus, we aim to solve directly the system of \cref{eq:RANSequations:div,eq:RANSequations:Mom}.
In order to close the system, BCs need to be defined for the various variables, \ie along $\Gamma_{i,o,t,b,B}$ presented in \cref{sch:coupling}. The knowledge of the potential field at those BCs is thus required. Because the time step and spatial discretizations of the HPC method are much large compared to the discretizations required by the RANS approach, interpolations are needed, both in time and space.

\paragraph{Spatial interpolation}~~\\ 
In practice and because we will make use of the same interpolation methods for the FD approach presented in \cref{sec:velocityDecomposition}, the spatial interpolation maps the external fields onto the internal mesh using \openfoam routines, based on a direct interpolation method, modified to efficiently handle large discretization differences. 

Assuming the knowledge of a field $q$  given on the external mesh at the cell centers, the procedure is as follows:
\begin{itemize}
    \item A projection onto the external mesh nodes (cell vertices) is first performed.
  \item For a spatial location of interest $\tI{x}$ (\ie a cell center/face center of the CFD mesh/boundary), we seek the cell belonging to the external mesh in which $\tI{x}$ is located.
  \item Afterwards, this cell is split into tetrahedron (note that the RANS model is applied on a 3D mesh of one cell width) and we identify which of them contains $\tI{x}$. Once done, we linearly reconstruct $q(\tI{x})$ using the values at the vertices of the tetrahedron.
\end{itemize}  
For all time steps at which the potential is available, the above method is applied, yielding the knowledge of the potential field projected onto the CFD mesh at each of these time steps.

\paragraph{Temporal interpolation}~~\\ 
Concerning the temporal interpolation, potential time instants framing the current time $t$ ($t_{n+1}<t\le t_n$) are identified and the corresponding spatial interpolations are used to approximate the field of interest at $t$. In practice, a linear temporal interpolation is used: $q(\tI{x},t)=  [(t_n-t)q(\tI{x},t_{n+1}) + (t-t_{n+1})q(\tI{x}, t_n)]/(t_n-t_{n+1}) $.

At this stage, approximations of the potential fields (velocity and pressure) are available on the CFD mesh at any given time, in particular at the boundaries of the CFD mesh. In the following these fields will be denoted $\vv_p$ and $p_p$.

Note that the procedure is also applied at the faces of CFD mesh boundaries, meaning that those fields can be used to enforce BCs.

\subsection{Boundary conditions}
\label{sec:DD:boundaryConditions}
With the knowledge of the potential fields (\ie velocity, pressure, \emph{etc}.), it is possible to implement different BCs. For example, the classical Dirichlet BC: 
\begin{equation}
    q(\tI{x}_f)=q_p(\tI{x}_f) \hspace{1cm} \forall f\in \Gamma
    \label{eq:Dirichlet}
\end{equation}
is implemented and available for both the pressure and the velocity. The use of this condition for both variables simultaneously is however too restrictive. Furthermore, enforcing \cref{eq:Dirichlet} on the velocity for a closed domain might lead to an unbalanced mass flux as the HPC model, the time and space interpolations and the different numerical treatments, may be prone to numerical errors leading to slightly non conservative fluxes across the boundaries.

For this reason, a variation of the \openfoam native \ofkw{inletOutlet} BC is employed for the velocity, denoted \ofkw{coupledVelocityInletOutlet}.
\begin{subequations}
    \begin{empheq}[left = \text{$\forall f \in \Gamma$}\empheqlbrace]{align}
     \text{if } \psi_f\le0  :\hspace{1cm}& \vv(\tI{x}_f)  = \vv_p(\tI{x}_f) \\ 
     \text{if } \psi_f>0  :\hspace{1cm}&  \grads{\vv} \cdot \tI{n}  = 0
    \end{empheq}
    \label{eq:velocityCoupledInletOutlet}
\end{subequations}
where $\psi_f$ is the phase flux at the given face $f$, positive for an outward flow. Thus, if the fluid flows outwards of the CFD domain, a null Neumann condition is applied. Reciprocally, if the fluid flows inwards, the condition reduces to a Dirichlet BC, where values from the potential model are enforced. Using this condition for the velocity with a Dirichlet BC for the pressure proved to return consistently accurate results with an increased stability.

\section{The Functional Decomposition (FD) approach}
\label{sec:velocityDecomposition}
The current implementation of the FD approach elaborates on the method introduced by \textcite{Beck:kim2005complementary}, in a similar fashion as the one presented in \textcite{zhang_multi-model_2018}. However, no transition zones are used here and thus the BCs are applied in a direct way at the outer perimeter of the internal domain. 
We wish to retain the domain reduction gain described in \cref{sec:domainCoupling}. Along with the fact that we keep a one-way coupling scheme, results are not expected to be in perfect agreement with the RANS method applied in an independent manner, partly for the same reasons that made the DD results differ from those of the standalone RANS simulation.
Thus, the method presented below is only applicable to cases where the viscous and turbulent effects do not perturb the far field flow in a significant way. Otherwise, stability issues should be expected at the BCs with a difficulty to drive the complementary values (and the turbulent eddy viscosity) to zero. We hypothesise that issues of this type led \textcite{zhang_multi-model_2018} to consider transition zones at the boundaries of the inner domain.
\subsection{Complementary RANS equations}
Following \citet{Beck:kim2005complementary}, the complementary counterpart $q^*$ of a given variable whose potential component is $q_p$ is defined so that:
\begin{equation}
    q_t=q_p + q^* 
    \label{eq:basicdecompositionAnyVariable} 
\end{equation}
Hereafter, for the sake of clarity, we use the subscript $t$ to denote total variables (this subscript should not be confused with the one used to denote the turbulent viscosity, $\mu_t$). $\vv_t$ and $p_t$ are thus the total velocity and pressure variables respectively, that are sought as solutions of the original NS or RANS equations. Note that $q_p$ can be obtained from any solver, and in the following, the only requirement is for $(\vv_p, p_p)$ to be solution of the Euler equations. 
Applied to the velocity, with a velocity deriving from a potential, $\vv_t$, this decomposition is a Helmholtz decomposition ($\vv_t=\rots \tI{a} + \grads \psi $ where $\tI{a}$ is a vector field and $\psi$ a scalar field). 
Thus, $\vv^*$ contains the rotational part of the total velocity. The Helmholtz decomposition is not unique and we focus on solving for $(\vv^*$, $p^*)$ that complement the potential velocity and pressure ($\tI{u}_p,p_p$) obtained from the FNPF-HPC model.

The velocity and pressure are replaced by their decomposition in \cref{eq:RANSequations:div,eq:RANSequations:Mom}. Simplifications are done stating that $(\vv_p, p_p)$ are solutions of the Euler equations. Afterwards, the RANS method presented in \cref{sec:RANS} is applied on the complementary velocity to yield the following equations, denoted the complementary RANS equations:
\begin{subequations}
    \begin{empheq}[left=\empheqlbrace]{align}
    & \divs \vv^*=0 \label{eq:cRANSequations:div}\\
    & 
    \begin{aligned}
    \dt{\rho \vv^* } 
          + \divs{ \rho\vv^*\crossop \vv^* } 
        & + \boxed{\divs{ \rho\vv_p\crossop \vv^* } }+  \boxed{\divs{ \rho\vv^*\crossop {\vv}_p } } \\
        & = - \grads{{p}^*}  
           - \divs \tII{T}_{\text{eff}}(\vv^*) 
    -\boxed{ \divs \tII{T}_t (\vv_p) } 
    \end{aligned}
    \label{eq:cRANSequations:Mom}
    \end{empheq}
\end{subequations}
The expression of $\tII{T}_{\text{eff}}(\vv^*)$ is given in \cref{eq:RANSequations:Mom}, using $\vv^*$ in place of $\vv$. The turbulent shear stress tensor applied on $\vv_p$ is given by:
\begin{equation}
    \tII{T}_t(\vv_p) =  \mu_t 
        (
            [ \grads \vv_p + (\grads \vv_p)^{T} ]  
            - \dfrac{2}{3} (\divs{\vv}_p)\id
        ) 
\end{equation}
Note that, because $\mu_t$ is not constant, the divergence of $\tII{T}_t$ does not cancel out, only the second part is null because of the divergence free property of $\vv_p$.

Given the previous equations, the form of the semi-discretized \cref{eq:HandAonU} remains intact. The differences are of course on how to compute diagonal and off-diagonal matrices of the momentum equation. More precisely, the different tensors in \cref{eq:HandAonU} are computed according to \cref{eq:cRANSequations:div,eq:cRANSequations:Mom} in place of \cref{eq:RANSequations:div,eq:RANSequations:Mom}. However given an equation, \openfoam offers numerical schemes to compute those terms. This implies that the pressure equation treatment does not need to be modified. Thus, the methodology described in \cref{sec:coupledVelocityPressure} is directly applied, and will not be repeated here. 

\subsection{Boundary conditions} \label{sec:VD:boundaryConditions}
In order to close the velocity-pressure problem, a set of boundary conditions has to be defined. While a DD coupling model has to enforce its outer boundaries according to the external solver values, it is not the case in the FD coupling framework. 
Deriving a Dirichlet BC -- \emph{i.e.} imposing a total velocity value $\vv_D$ such that $\vv_t = \vv_D$ at the BC -- in this framework yields:
\begin{equation}
    \vv^* = \vv_D-\vv_p
    \label{eq:dirichelOnUstart}
\end{equation}
In particular, in the case of a no-slip condition (\ie $\vv_D=\tI{0}$, sea bottom or non moving body BC, \emph{e.g.} $\Gamma_B$ on \cref{sch:coupling}), we get:
\begin{equation}
    \vv^* = -\vv_p \text{\ \ at \ $\Gamma_B$} 
    \label{eq:velocityCouplingNoSlip}
\end{equation}
At the outer boundary ($\Gamma_t$, $\Gamma_b$, $\Gamma_i$ and $\Gamma_o$ on \cref{sch:coupling}), the same Dirichlet condition can be applied. It is assumed that the effects described by the complementary equations are restricted to the body vicinity and thus, the condition is reduced to:
\begin{equation}
    \vv^*=\tI{0}
    \label{eq:velocityCoupledDirichletOuterBnd}
\end{equation}
A Neumann BC on which the imposed spatial derivative of the total velocity is imposed as $\left.\dn{\vv}\right|_N$ would reduce to:
\begin{equation}
    \dn{\vv^*} = \left.\dn{\vv}\right|_N - \dn{\vv_p}
    \label{eq:velocityCouplingNeumann}
\end{equation}

Because of the one-way nature of the coupling scheme, stability problems are expected when imposing \cref{eq:velocityCoupledDirichletOuterBnd} at the outer boundaries of the RANS domain: the feedback of the turbulent and viscous part to the far-field flow cannot be exactly zero. Thus, in order to counteract this, a mixed Neumann-Dirichlet BC will be used, in the same manner as in \cref{sec:domainCoupling}, derived here for a complementary variable:
\begin{subequations}
    \begin{empheq}[left = \text{$\forall f \in \Gamma$}\empheqlbrace]{align}
     \text{if } \psi_f\le0  :\hspace{1cm}& \vv^*(\tI{x}_f)  = \tI{0} \\ 
     \text{if } \psi_f>0  :\hspace{1cm}&  \grads{\vv^*} \cdot \tI{n}  = \tI{0}
    \end{empheq}
    \label{eq:velocityCoupledInletOutletUc}
\end{subequations}
Note that multiple face fluxes are available in this context, namely the potential one $\psi_p$, the complementary one $\psi^*$ and the total one $\psi_t$. In the present implementation, the user is free to select any one of them. Selecting $\psi_t$ or $\psi_p$ at the outer boundaries of the domain ($\Gamma_{i,o,t,b}$ on \cref{sch:coupling}) should yield the same results, given that we expect the complementary effects to vanish at those boundaries. In practice, it was indeed found no difference when selecting either the potential face flux or the total one. Selecting $\psi^*$ however, would imply that a null Dirichlet BC is enforced most of the time, allowing for a flow only when non null complementary velocities are found in the boundary vicinity.

\subsection{Expected sources of discrepancies between DD and FD coupling methods}\label{sec:expectedDiscrepancies}
While in theory the FD and DD approaches are mathematically equivalent, some sources of discrepancies are expected due to numerical treatments.

\paragraph{Residual threshold}~~\\ 
When inverting the matrix of the Poisson \cref{eq:pEqn} with any of the available matrix solvers, a target residual (tolerance) needs to be specified. This tolerance is compared with the solution residuals scaled 
by a value representative of the magnitude of the field that is solved for.

However, the complementary fields magnitudes might differ from the total fields magnitudes. Thus, if one wants to obtain the same precision on the \emph{total} fields, the target residuals might have to be modified. In particular, it was found that the complementary pressure field is more than one order of magnitude lower than the total dynamic pressure field. For this reason, the effect of the  tolerances selection is analysed and the results will be discussed in \cref{sec:sensitivityResultTolerance}. 

\paragraph{Nonlinear discretization schemes}~~\\ 
In order to derive \cref{eq:cRANSequations:div,eq:cRANSequations:Mom} from \cref{eq:RANSequations:div,eq:RANSequations:Mom}, we largely took advantage of the linear properties of the divergence and gradient operators. However, some of the available discretization schemes can be nonlinear with $\vv$. For example, the upwind divergence scheme, classically applied for the discretization of the velocity advection, is not fully linear with $\vv$, \ie, numerically:
\begin{equation}
    \divs \rho \vv_t \crossop \vv^*  = \divs \rho (\vv_p+\vv^*) \crossop \vv^* \ne \divs \rho \vv_p \crossop \vv^*+ \divs \rho \vv^* \crossop \vv^*
    \label{eq:aggregatedVsSeparated}
\end{equation}
even though the equality should be mathematically verified. The influence on the resolution of this term discretized following a 'separated' approach (right part of \cref{eq:aggregatedVsSeparated}) through two different upwind schemes  compared with an 'aggregated' manner (left part of \cref{eq:aggregatedVsSeparated}) was conducted, showing that the aggregation has a stabilization effect as long as multiple \ofkw{PIMPLE} iterations are performed.

Note however that some terms were simplified invoking the Euler equation for the potential variables. Thus, $\divs \rho \vv_p  \crossop \vv_p$ is not available anymore in \cref{eq:cRANSequations:Mom}: $\divs \rho \vv^*  \crossop \vv_p$ remains solely and its discretization might lead to discrepancies when comparing with the DD approach.

\section{Models comparison: coupling validation and analysis}
\label{sec:resultsModelsComp}

In order to validate and compare the results obtained \emph{via} the FNPF-HPC model and the two decomposition approaches, an experimental wave-structure interaction case, detailed in \cref{sec:casedescription}, was selected. Computed loads on the structure, as well as vorticity fields, will then be compared with a standalone CFD simulation of the same case done with \openfoam and denoted \ofkw{waveFoam}, referring to the name of the used solver \parencite{jacobsenFuhrmanFredsoe2012}.

\subsection{Case description} \label{sec:casedescription}

\subsubsection{Geometry}
\begin{figure}[htb!]
    \centering
    \includegraphics{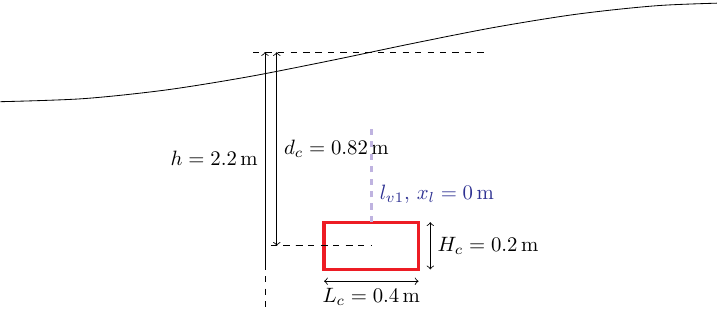}
    \caption{Schematic representation of the selected case. Note that the sea bottom is not represented, and everything is at scale, considering a wavelength $\lambda=\SI{6.15}{\meter}$ (half of it represented here) and a wave steepness $H/\lambda=7.0\%$.}
    \label{sch:cylinderVenugSch3}
\end{figure}

A fully immersed horizontal cylinder of rectangular cross-section is selected, reproducing the experimental studies of \textcite{arai1995forces,venugopal_hydrodynamic_2002}. An aspect ratio height over length of the rectangle $H_c/L_c=1/2$ is selected here. The relative submergence depth of the center of the rectangle with respect to the Still Water Level (SWL) is set as $d_c/H_c=4.1$ and the still water depth is set to $h/d_c=2.68$. A sketch of the case with the main dimensions is given in \cref{sch:cylinderVenugSch3}. Note that spatial profiles of flow variables will be sampled in the following along the vertical dashed line {\color{Blue} $l_{v1}$}, from the body upper wall to the CFD mesh top boundary ($\Gamma_t$ in \cref{sch:coupling}).

\subsubsection{Turbulence modeling}\label{sec:turbulenceModels}
The selected model is the \kos \parencite{menter_two-equation_1994}, for its capability to capture both wall vicinity and larger scale flows. Because no model is available in \openfoam (v1712) for multiphase flows, the correction made available by \textcite{devolder_application_2017,devolder_performance_2018} is used instead. Turbulence models were however shown to be formally unstable under potential waves by \textcite{larsen2018over}, who suggested a limitation based on the magnitude of the rotation rate tensor. This limiter however proved to be intrusive  in the vortex shedding zones on the selected case, and is therefore not used here. 
Futher investigations are currently conducted with the aim to fully understand this shortcoming and suggest a correction.

\subsubsection{Incoming waves}\label{sec:incomingWaves}
In this study, incoming waves are selected as regular with fixed period $T=\SI{2}{\second}$. %
Different wave heights will be simulated in \cref{sec:resultsVsLiterature}, though most of the simulations of this section consider a fixed wave steepness $H/\lambda = 3.5\%$.  

Because of the choice of the turbulence model, see \cref{sec:turbulenceModels}, the turbulence growth is damped at the air-water interface but still arises in the bulk of the fluid domain. Energy loss is thus expected when \openfoam is used to propagate water waves over distances of a few wavelengths, contrarily to what is observed with the FNPF-HPC solver. In summary, within either the DD or FD coupling approach, the cylinder might not be subjected to the same wave and local flow field as in the standalone CFD approach, and the former is expected to be more accurate regarding the quality of the propagated wave field at the body location. 
\subsubsection{Boundary conditions}
\renewcommand{\arraystretch}{1.3}
\begin{table}[tbp!]
    \newcommand{\mr}[2]{\multirow{#1}{*}{#2}}
    \centering
\scriptsize{
    \begin{tabular}{|c|c|c|c|c|c|}
        \hline
        boundary                   & model & $\vv$; $\vv^*$(\si{\meter\per\second}) & $p-\rho g h$; $p^*$                 & $k$ (\si{\kunit})                   & $\omega$ (\si{\per\second}) \\ \hline
        \mr{2}{$\Gamma_{i,t,o,b}$} & DD    & cIO                                    & cfV                                 & \mr{2}{fV:$10^{-10}$}              & \mr{2}{fV: 100} \\ \cline{2-4}
        & FD    & IO:$\tI{0}$                            & fV:$0$                              &                                    & \\ \cline{1-1} \cline{2-6}
        \mr{2}{$\Gamma_{B}$ }      & DD    & fV:$\tI{0}$                            & \mr{2}{\texttt{fixedFluxPressure}:0} & \mr{2}{\texttt{ kqRWallFunction }} & \mr{2}{\texttt{omegaWallFunction}} \\ \cline{2-3}
                                   & FD    & -cfV                                   &                                     &                                    & \\ \cline{0-5}
        \hline
     \end{tabular}
   }
   \caption{Table of the tested sets of boundary conditions (see \cref{sch:coupling} for the geometrical definition of the BC). IO stands for \ofkw{inletOutlet}, fV for \ofkw{fixedValue}. A prefix ``c'' is added when their coupled counterparts are used. For details about these boundary conditions, see \cref{sec:DD:boundaryConditions,sec:VD:boundaryConditions}.}
     \label{tab:testedBcSetsDomainCoupling}
 \end{table}
 \Cref{tab:testedBcSetsDomainCoupling} lists the types of the boundary conditions used in the coupled models. A wall model is used for the body boundary condition of the turbulence variables that allow to have an $y^+$ located either in the low Reynolds zone or in the log-profile region. In the case with $H/\lambda=3.5\%$ the maximum (over time and space) encountered value of $y^+$  is about 6, granting trust on the proper resolution of the boundary layer flow. 

Values of potential variables at the coupled boundaries are interpolated from the obtained fields with the FNPF-HPC model that was run \emph{a priori}. Note that results obtained independently with this model (thus corresponding to a FNPF simulation) will also be shown in the following comparisons. 

\label{sec:largeOmegaNew}
Note that we enforced a high value of the dissipation rate $\omega$ (\cf \cref{tab:testedBcSetsDomainCoupling}) at the outer boundaries, because otherwise, the expected rise of turbulent viscosity in their vicinity yields nonphysical peaks due to the difficulty to respect a null TKE value. This high dissipation rate value, though not physically consistent, is numerically beneficial in terms of stability and it does not modify in a significant manner the wall vicinity flow (as shown later in \cref{sec:largeOmega}). Furthermore, it is consistent with the underlying hypothesis of fading away turbulence when approaching the outer boundary of the local domain.    

\subsubsection{Numerical parameters and meshes} \label{sec:numericalparameters}
With the aim of accurately comparing the coupling methods with the standalone \ofkw{waveFoam} simulation, most of the numerical parameters were maintained identical. Among them, the gradient operator is discretized with a least-square limited method, and the advection of the velocity with a limited linear scheme. %
A linear scheme without any limiter is also used for all other divergence terms, excepts the convection of turbulent variables for which a upwind scheme is used instead. Finally, the Laplacian operator is discretized with a linear method, with a correction for non-orthogonality of the mesh.

Note that the FD method requires the \ofkw{PIMPLE} loop to be active with an exit residual target specified in order to yield stable results. 

The mesh itself also inherits most of the features from the \ofkw{waveFoam} mesh: the wall vicinity is discretized such that the boundary layer can be accurately resolved, a refined discretization is used close to the body corners. The overall body vicinity discretization is also respected, with the use of square cells of dimension $dx=0.056\si{\meter}$. A mesh independence study (not shown here for brevity) was performed to select this value. 

When the flow on the whole domain is computed (\emph{i.e.} HPC and \ofkw{waveFoam} computations), we select the mesh span as approximately $8\lambda$  with both the generating and absorbing relaxation zones being of length $2\lambda$. The cylinder is located at the center of the domain.

The time step is fixed at a value of $dt=T/4000=\SI{5e-4}{\second}$ in both coupled models as well as in the independent RANS approach. 

For the HPC method, a dual mesh method (boundary fitted overlapping grid) is used, as presented in \textcite{robaux_development_2021-1}. The background mesh, used to propagate the incident and reflected waves is discretized with $dx=\lambda/60$, and the time step is also fixed, at a value controlled by the Courant-Friedrichs-Lewy (CFL) number based of the phase velocity of the waves. This CFL is fixed at $2$, leading to a time step of $dt=T/30=\SI{0.067}{\second}$.

\subsection{Temporal loads series} \label{sec:temporalloads}
\begin{figure}[htb!]
    \centering
    \includegraphics{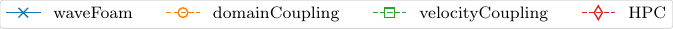}

    \sfig{plottemporalLoadsHPCOFvCdCwF_nomarkers}[0.59]
    \sfig{plottemporalLoadsHPCOFvCdCwFZoomed}[0.39]
    \caption{Temporal series of the loads obtained in the case $T=2$~s and $H/\lambda = 3.5\%$ with the waveFoam solver, the DD (domainCoupling) method, the FD (velocityCoupling) method and the standalone HPC code. Note that the coupled approaches are started at $t/T=12$ (\ie at $t=24$~s) as explained in \cref{sec:hotstartcapabilities}. }
    \label{fig:plottemporalLoadsHPCOFvCdCwF}
\end{figure}

At any time instant, the force applied on the cylinder (per unit width in the transverse direction) is calculated by integrating the stress along the body boundary, and decomposed into a horizontal component $(f_x)$ and a vertical component $(f_z)$.
The temporal load series obtained with the four presented models are depicted in \cref{fig:plottemporalLoadsHPCOFvCdCwF} for the case $T=2$~s and $H/\lambda = 3.5\%$. While a relative agreement can be found between all four models, it is noticeable that the coupled methods are able, making use of the HPC results, to recover the loads obtained with waveFoam, especially the horizontal one. We indeed remark a good agreement of the loads after a few periods of evolution. Note that the coupled models are started at a time when the HPC method already yields stable and periodic results here at $t=12T=24$~s. Eventually, the right panels focus on a one wave period time range $[18T,19T]$ after a duration of $6T$ has been simulated with the coupled methods, while a duration of $18T$ was necessary with HPC and waveFoam used as standalone codes.

We note the two coupled models are in very good agreement concerning both load components. This fact tends to validate the correct implementation of both models as they both work with equivalent boundary conditions. 

However, the vertical loads from coupled models compare better with HPC than with waveFoam, denoting a lower impact on this force component of the turbulent and rotational effects. It will be shown in \cref{sec:localFieldsDescriptions} that the imposed pressure at the top boundary conditions differs from the waveFoam pressure and that this discrepancy propagates down to the cylinder wall.
As stated earlier, the incoming waves predicted by HPC and waveFoam slightly differ in height at the body location, which might also contribute to the obtained difference in terms of vertical load.

\begin{figure}[htb!]
    \centering
    \includegraphics{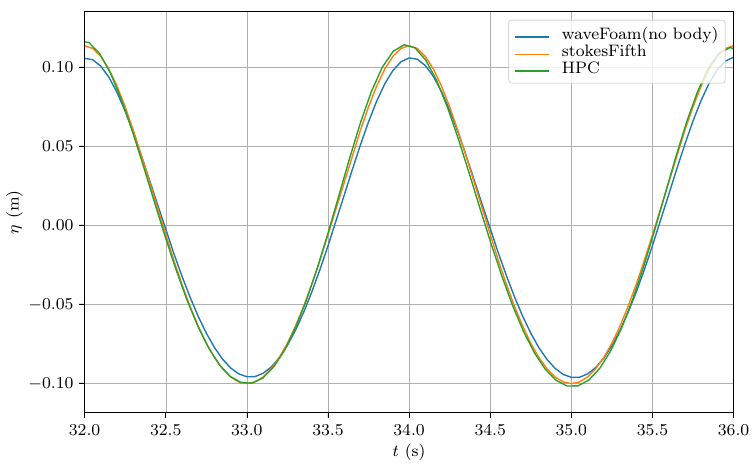}

    \caption{Temporal series of the free surface elevation for the case $H/\lambda=3.5\%$ at the abscissa of the body from a standalone waveFoam simulation in the absence of the body, compared with the 5th order Stokes solution (stokesFifth curve) and the HPC simulation used in the two coupled approaches (\emph{i.e.} with the presence of the body).}
    \label{fig:plottemporaleta}
\end{figure}

This difference in terms of incoming waves is further evidenced by a standalone computation with waveFoam, done without the body, and we compare in \cref{fig:plottemporaleta} its results with those of the HPC model at this same location as well as the prediction from a Stokes 5th order wave model. We clearly note that a propagation error develops within the domain with the standalone CFD simulation, leading to a slightly lower wave amplitude at the body location. Note that the waveFoam simulation uses the same numerical parameters (discretization schemes, time step, \emph{etc.}) as the ones used in the coupled approaches, so aimed at correctly representing the flow in the body vicinity. Given the high sensitivity of the wave propagation to those numerical parameters \citep[see \eg][]{larsen_performance_2019}  more efforts could be devoted to obtain a more accurate propagation phase of the incoming waves. Note also that the HPC result for free surface elevation time series in \cref{fig:plottemporaleta} is obtained from a simulation considering the body presence, which might explain the very small discrepancies with the theoretical wave profile. For further assessment about the performance of the HPC model in cases of pure wave propagation, the reader is referred to \citet{robaux_development_2021-1}.

\subsection{Vorticity fields}
\begin{figure}[htb!]
    \centering
    \begin{subfigure}[c]{0.48\textwidth}
            \centering
            \includegraphics[width=0.99\textwidth]{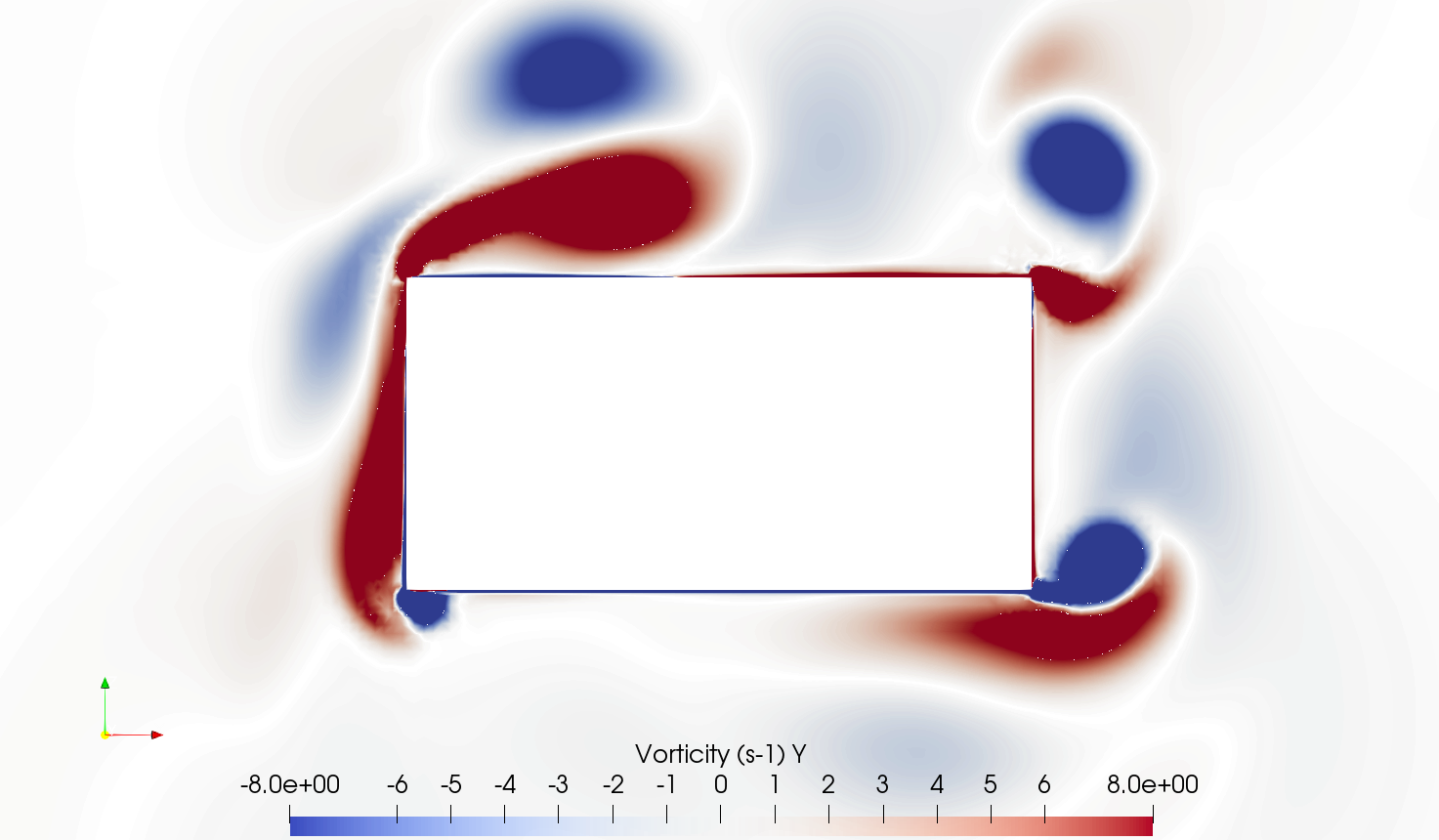}
            \caption{waveFoam}
            \label{fig:sub:WFvorticity}
    \end{subfigure}%
    \begin{subfigure}[c]{0.48\textwidth}
            \centering
            \includegraphics[width=0.99\textwidth]{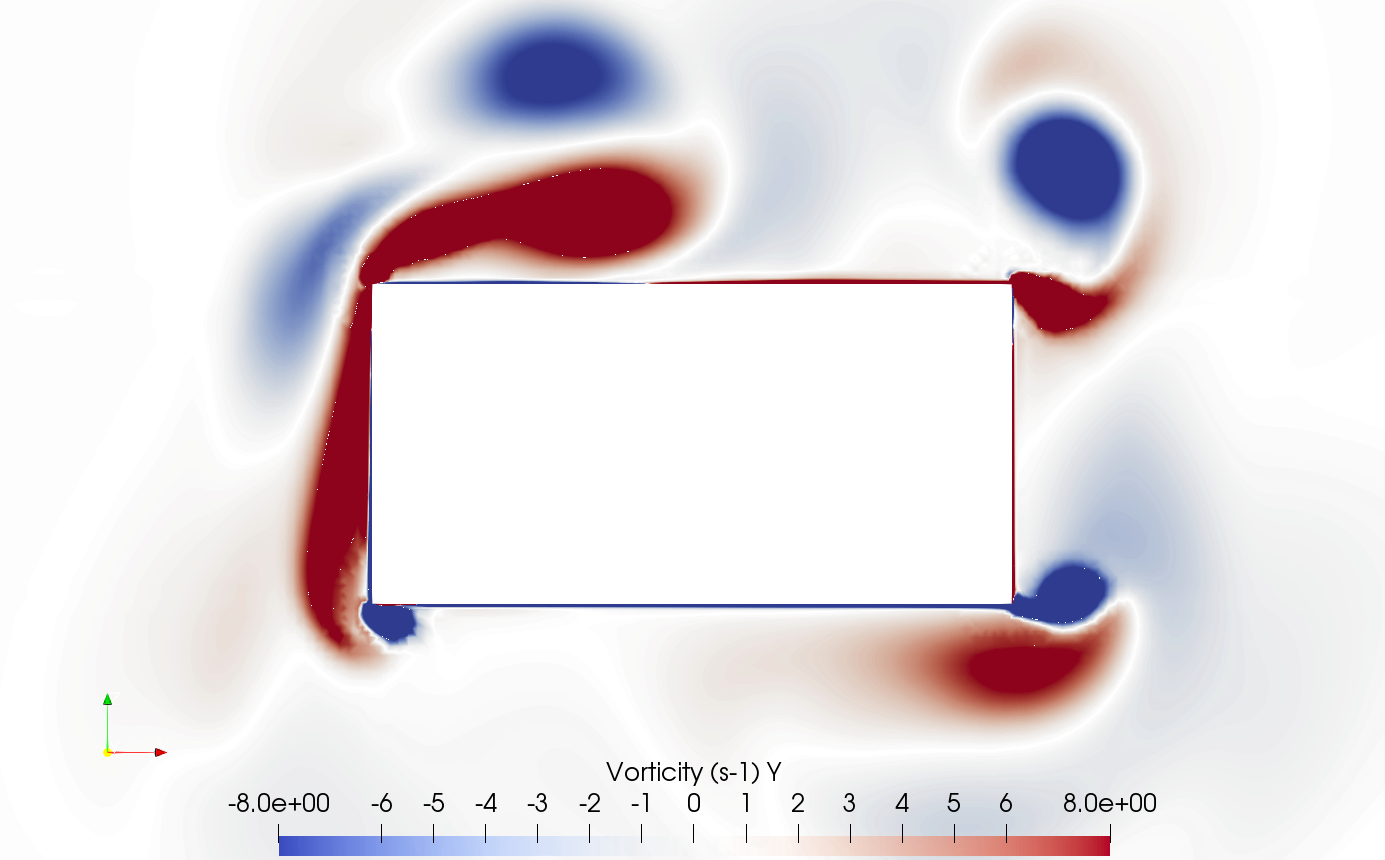}
            \caption{domain coupling}
            \label{fig:sub:DDvorticity}
   \end{subfigure}%

    \begin{subfigure}[c]{0.48\textwidth}
            \centering
            \includegraphics[width=0.99\textwidth]{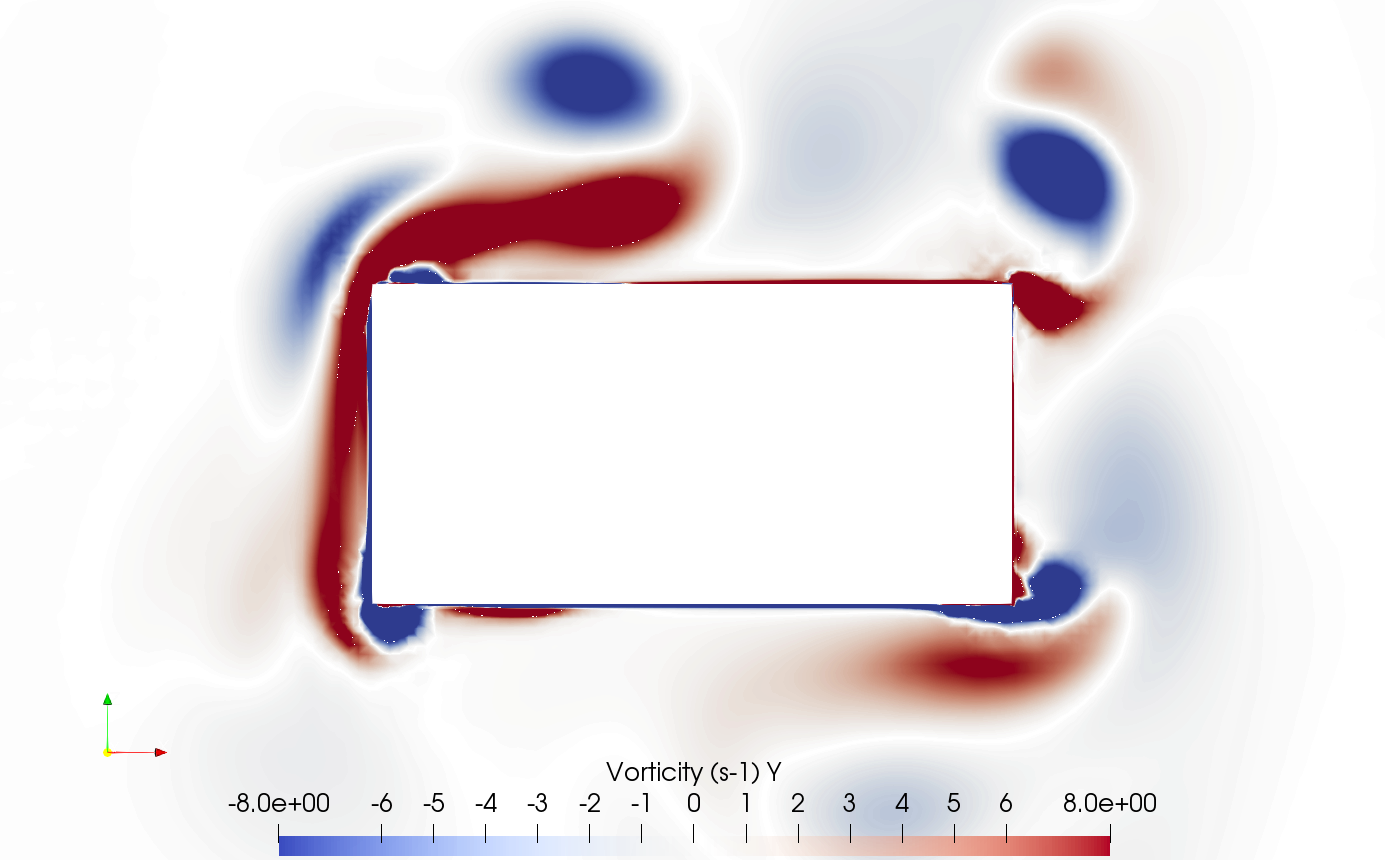}
            \caption{functional decomposition}
            \label{fig:sub:VDvorticity}
        \end{subfigure}%

    \caption{Comparison of the vorticity fields predicted by (a) waveFoam and the coupled models ((b) DD method; (c) FD method) at $t/T=18$, \ie  once a periodic behavior is established (case $T=2$~s and $H/\lambda = 3.5\%$).}
    \label{fig:vorticityFields}
\end{figure}

The FNPF-HPC model, used to impose the boundary conditions of the coupled approaches, neglects flow vorticity. The underlying hypothesis is that the extent of the CFD mesh covers the entire region where the vorticity cannot be neglected. On \cref{fig:vorticityFields}, the vorticity fields from the coupled approaches are shown together with the vorticity field predicted by the \ofkw{waveFoam} standalone computation. 

A good visual agreement can be denoted, emphasizing the capabilities of both coupled approaches to recover vorticity effects, despite the enforcement of BC with null vorticity. Note also that this figure tends to validate the underlying assumption that the CFD mesh covers the entire region where vorticity remains significant.

\subsection{Local field descriptions}\label{sec:localFieldsDescriptions}

In order to compare models results in a more quantitative manner, flow variables are sampled over a vertical line starting at the center of the upper wall of the body section and ending at the upper boundary of the local mesh around the body (\textcolor{Blue}{$l_{v1}$} on \cref{sch:cylinderVenugSch3}) at 40 different times during a given wave period. 

\begin{figure}[htb!]
    \centering
    \includegraphics{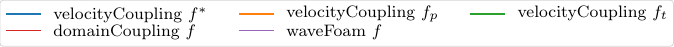}

    \sfig{plotOverLineMultiCasevCvsdCU}
    \sfig{plotOverLineMultiCasevCvsdCU2}
    \sfig{plotOverLineMultiCasevCvsdCP}
    \caption[velocity Coupling local results]{Comparison of the complementary fields, the potential ones and the resulting total ones from the FD approach (velocityCoupling) with waveFoam and the DD solver (domainCoupling) at a particular time $t/T=20$ sampled along a vertical segment on top of the body upper face, denoted \textcolor{Blue}{$l_{v1}$} on \cref{sch:cylinderVenugSch3} (case $T=2$~s and $H/\lambda = 3.5\%$).}
    \label{fig:plotOverLineMultiCasevCvsdCU}
\end{figure}

\paragraph{Comparisons at a given time}~~\\ 
An example of the obtained fields is given in \cref{fig:plotOverLineMultiCasevCvsdCU} at a particular time for the velocity components and pressure. Note that the potential fields are also included, which serve for imposing both the boundary conditions in the two coupled approaches, and the potential fields in the FD approach. For the FD method, the three available fields are represented, namely the potential components, the complementary components and their sum, \ie the total fields. 

It can be noted that the pressure computed from HPC is slightly different from the one predicted by waveFoam, from the body upper wall ($z=\SI{-0.72}{\meter}$) to the top boundary ($z=\SI{-0.3}{\meter}$). This discrepancy, enforced in the coupled approaches at the top boundary, propagates down the body walls. 
It is thought to be the reason of the vertical loads discrepancies observed in \cref{fig:plottemporalLoadsHPCOFvCdCwF}. %
It is however possible to note that the near wall evolution of the pressure from the two coupling methods is similar to the one obtained with waveFoam (and thus distinct from the HPC pressure), maintaining the shift due to the HPC pressure further away. This is noticeable for example by denoting that the complementary pressure is almost null (\ie $p=p_p$) away from the body, but exhibits a small value in the wall vicinity.  
Finally, we observe that while the complementary velocity, which ``corrects'' the potential values, is of large magnitude compared to the potential one, the complementary pressure is very small. 

\paragraph{Comparisons over a wave period}~~\\ 
Hereafter and for the rest of this study, only the total fields -- \emph{i.e.} the sum of the potential and complementary fields -- obtained with the FD method will be shown for clarity reasons. 

\begin{figure}[htbp!]
    \newcommand{\vs}{\vspace{-0.12cm}}
    \centering
    \vs
    \vs
    \vs
    \includegraphics{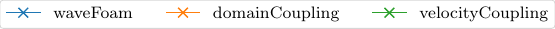}

    \vs
        \begin{subfigure}[c]{0.5\textwidth}
                \includegraphics{%
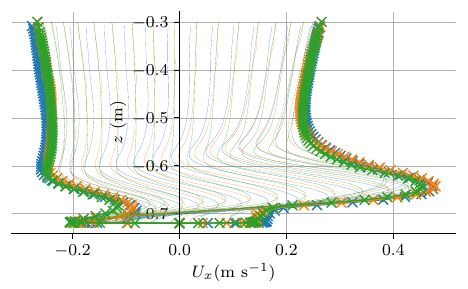} 
     \caption{ $\bar { t } <0.5$      }
        \end{subfigure}%
        \begin{subfigure}[c]{0.5\textwidth}
                \includegraphics{%
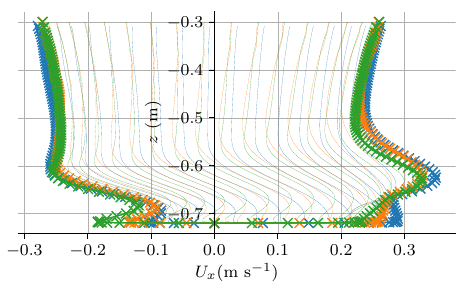} 
     \caption{ $\bar { t } \ge 0.5$ }
        \end{subfigure}%

    \vs
        \begin{subfigure}[c]{0.5\textwidth}
                \includegraphics{%
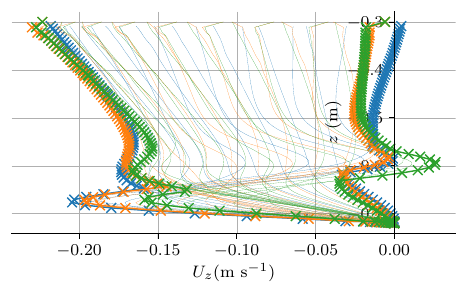} 
     \caption{ $\bar { t } <0.5$      }
        \end{subfigure}%
        \begin{subfigure}[c]{0.5\textwidth}
                \includegraphics{%
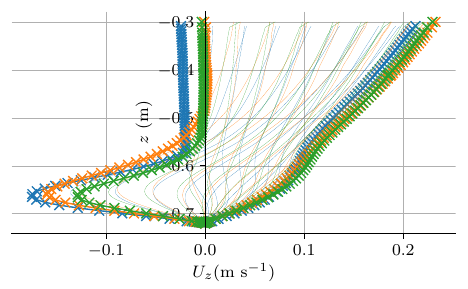} 
     \caption{ $\bar { t } \ge 0.5$  }
        \end{subfigure}%

    \vs
        \begin{subfigure}[c]{0.5\textwidth}
                \includegraphics{%
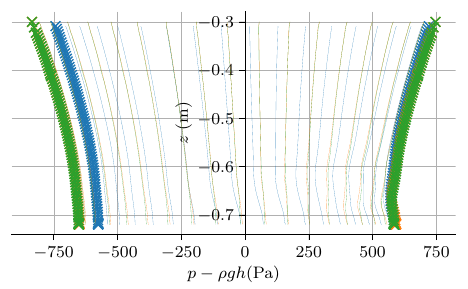} 
     \caption{ $\bar { t } <0.5$      }
        \end{subfigure}%
        \begin{subfigure}[c]{0.5\textwidth}
                \includegraphics{%
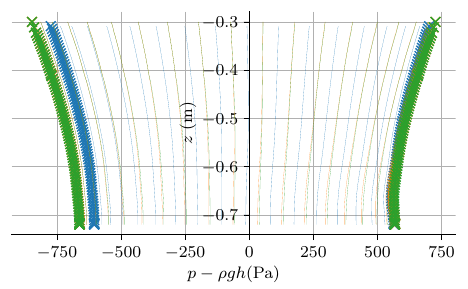} 
     \caption{ $\bar { t } \ge 0.5$  }
        \end{subfigure}%

    \vs
        \begin{subfigure}[c]{0.5\textwidth}
                \includegraphics{%
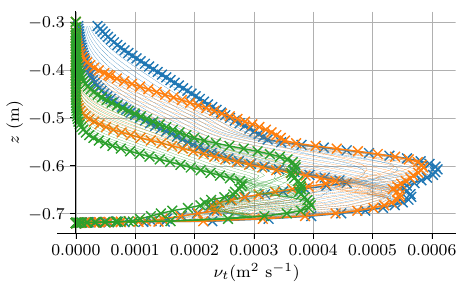} 
     \caption{ $\bar { t } <0.5$      }
        \end{subfigure}%
        \begin{subfigure}[c]{0.5\textwidth}
                \includegraphics{%
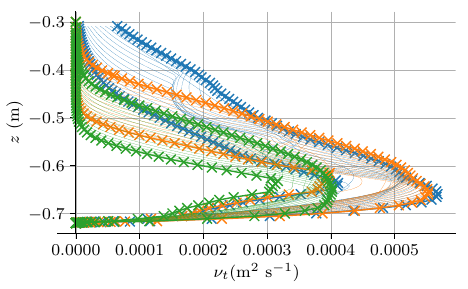} 
     \caption{ $\bar { t } \ge 0.5$  }
        \end{subfigure}%

    \vs
    \caption{Total fields (horizontal velocity, vertical velocity, dynamic pressure and turbulent kinematic viscosity $\nu_t=\mu_t/\rho$) from waveFoam, DD (domainCoupling) and FD (velocityCoupling), sampled over a vertical line ($x_l=\SI{0}{\meter}$, $z=\ $\SIrange{-0.72}{-0.3}{\meter}, \textcolor{Blue}{$l_{v1}$} on \cref{sch:cylinderVenugSch3}) at $40$ time instants separated into two half wave periods (left and right panels).} 
    \label{fig:plotBoundaryLayer2Meshes1PeriodvCvsWF0}
\end{figure}

All selected 40 time steps, as well as the obtained envelopes are shown in \cref{fig:plotBoundaryLayer2Meshes1PeriodvCvsWF0}. Left panels correspond to the first half wave period, and right panels to the second half period. 

We note that both coupling approaches recover the expected horizontal velocity profile as well as its envelopes in an accurate manner. Time step curves discrepancies can be attributed to a small phase shift, that we know will happen during the wave propagation process, mainly with the waveFoam approach. 

The pressure profiles are also recovered even though the added pressure by the coupled approaches were shown to be of small magnitude compared to the HPC results. The obtained magnitudes are thus not very different from the HPC potential pressure profiles (not shown here).

The DD approach seems to accurately capture all profile and envelopes, at least in the body vicinity. Note that the turbulence is neglected at the top boundary of the local mesh ($z=-0.3$~m) in the coupled approaches, but is not found to be null at that location by waveFoam. As presented in \cref{sec:largeOmegaNew} and \cref{tab:testedBcSetsDomainCoupling}, a large value of $\omega$ is enforced at the top boundary condition. This acts as a transition zone to smoothly impose the hypothesis of null turbulent viscosity at the BC.\label{sec:largeOmega}

Results of the FD approach on the vertical velocity as well as on the turbulent viscosity exhibit more discrepancies than the DD approach with the independent RANS simulation. This is thought to be arising from the possible sources of discrepancy discussed previously.

\subsection{Parameters investigations}
\subsubsection{Assessment of hot start capabilities of the DD approach } \label{sec:hotstartcapabilities}

\begin{figure}[htb!]
    \centering
    \includegraphics{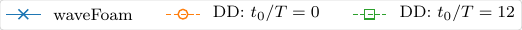}

    \sfig{plottemporalLoadsHPCOFCompareStartTime_nomarkers}[0.59]
    \sfig[Zoom]{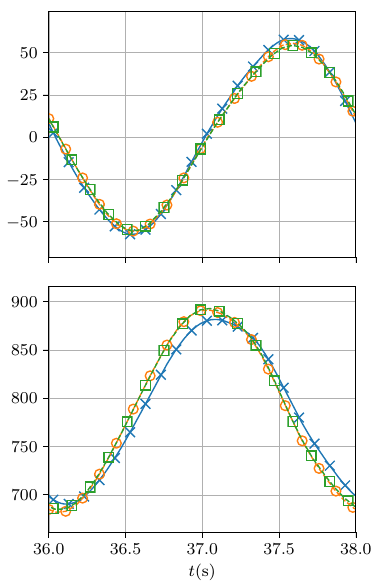}[0.39]
    \caption{Temporal series of the loads when performing a start of the DD coupling method at $t_0/T=0$ and a ``hot start'' a time $t_0/T=12$, compared with waveFoam standalone simulation. See text for details.} 
    \label{fig:plottemporalLoadsHPCFOFCompareStartTime}
\end{figure}

\Cref{fig:plottemporalLoadsHPCFOFCompareStartTime} shows the time series of the loads obtained with waveFoam and the DD approach with two different starting times. The first one is started at $t_0/T=0$, meaning that for a significant duration ($t/T\in\sim[0,5]$), the CFD solver receives mostly null values at its boundary conditions: the wave did not propagate up to the body location yet. If the maximum local CFL number were used to control the time step, this first phase would be relatively cheap in terms of computational cost. Afterwards, a second phase occurs on which the CFD code solves the problem with boundary conditions that did not reach a periodic behavior yet (a CFL control would not decrease the CPU cost of this phase): if one is not interested in the transient behavior, this phase could also be skipped. Only later, at $t/T\approx 10$, a periodic and converged wave field is expected to be imposed at the boundaries on the local domain. 

For this reason, a second run of the DD coupled model is done and started at $t_0/T=12$ (denoted ``hot start''), with the potential fields mapped on the CFD local mesh as starting conditions. With this approach, the expected number of simulated wave periods required to reach a periodic state is lowered. 
In practice, we found that the duration of the unphysical transient that needs to be simulated to achieve periodic results is approximately $6T$ when this later starting time is used (\eg $t_0/T=12$ here), while a duration of at least $15T$ was mandatory when starting the DD case at $t_0/T=0$.

The comparison of loads after those $6T$ presented in the right panels of \cref{fig:plottemporalLoadsHPCFOFCompareStartTime} shows a good agreement, supporting the assumption that running the DD simulation from $t_0/T=0$ increases the CPU time without any added value.

\begin{figure}[!htb]
        \centering
    \includegraphics{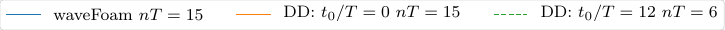}

        \sfig[]    {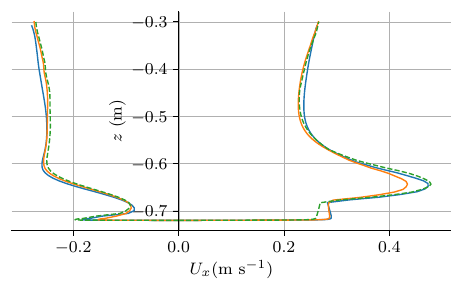}
        \sfig[]    {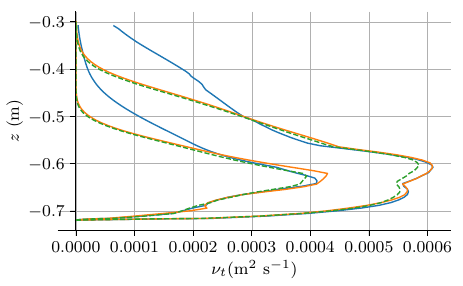}

        \caption{Horizontal velocity (\cref{fig:sub:plotBoundaryLayer2Meshes1PeriodU0hotstart0}) and turbulent kinematic viscosity (\cref{fig:sub:plotBoundaryLayer2Meshes1PeriodNuthotstart0}) profiles envelopes along the transect {\color{Blue} $l_{v1}$} (see \cref{sch:cylinderVenugSch3}), computed from different 40 time steps values per wave period. Different starting times ($t_0$) are used, fields are therefore compared after $nT$ simulated periods. 
        }
        \label{fig:plotBoundaryLayer2Meshes1PeriodU0hotstart0}
\end{figure}

\Cref{fig:plotBoundaryLayer2Meshes1PeriodU0hotstart0} further confirms this conclusion, by showing a good comparison between the local fields obtained with the DD approach with two different starting times.

\subsubsection{Sensitivity to the mesh horizontal extent of the DD approach} \label{sec:mesh_breatdh}

\begin{figure}[htb!]
    \centering
        \includegraphics{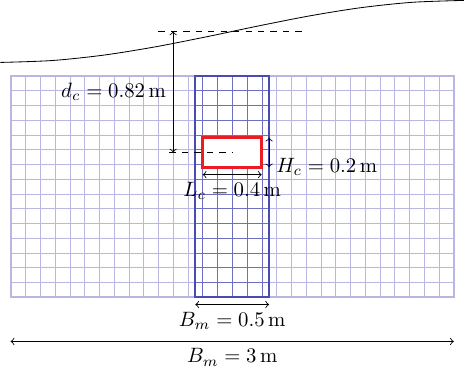}
        \caption{Different CFD mesh horizontal spans parameterized by $B_m$. Sketch at scale.}
    \label{fig:schematicBm}
\end{figure}

All the previously presented computations were done with a local CFD mesh horizontal extent of $B_m = 7.5 L_c = 3\si{\meter}$. Because a reduced extent of the local mesh would be computationally faster, a convergence study is performed on this parameter, varying $B_m$ from a very low value of $1.25 L_c = 0.5\si{\meter}$ to the above value $7.5 L_c$ (see \cref{fig:schematicBm}).  Hence, the goals here are i) to validate that the current mesh extent yields converged results, and ii) to investigate on its lower limits to evaluate how much computational power could be saved. 

\begin{figure}[htb!]
    \centering
        \includegraphics{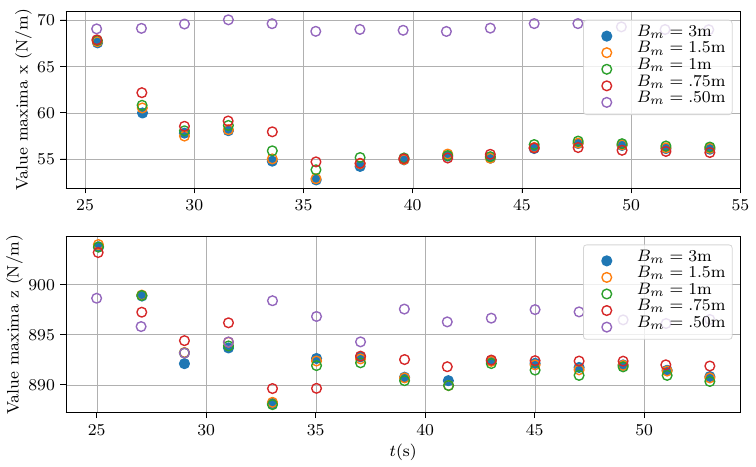}
        \caption{Maximum values of obtained loads for different CFD mesh horizontal extents predicted with the DD approach.}
    \label{fig:plottemporalAmplitudeErrorbreadthErrors}
\end{figure}

The maximum values of the loads computed with the DD method are shown on \cref{fig:plottemporalAmplitudeErrorbreadthErrors}. We note that, while some differences can be seen, they are of small amplitudes as soon as $B_m \ge \SI{0.75}{\meter}$, which already yields valuable results. 
Further reducing the mesh width (\ie $B_m=\SI{0.5}{\meter}$) leads to large discrepancies. Note that the horizontal width of the body is $L_c=\SI{0.4}{\meter}$, thus, a $\SI{0.5}{\meter}$ domain extent means that the extent of the CFD domain, past the body vertical sides, is only of $\SI{0.05}{\meter} =L_c/8$. 

Let us define this horizontal meshed length on each side of the cylinder $\delta l_m = (B_m -L_c)/2$ and $A_m$ the horizontal excursion of a fluid particle at $z=-d_c$ (which will also be used to define the Keulegan-Carpenter number, see \cref{sec:KCdefinition}). Thus, relative to $A_m$, the meshed part on each side is $\delta l_m / A_m\approx 1/6$ with the shortest domain. With a mesh extent of $B_m=\SI{0.75}{\meter}$, the same parameter is $1/1.7$ and reaches $1$ for $B_m=\SI{1}{\meter}$.
This means that using a domain with an extent of 1/1.7 to 1 times the amplitude of motion of a particle on each side of the cylinder is sufficient to capture most of the turbulent and vorticity effects in the vicinity of the body. To remain conservative and because the computation cost is not much higher (most cells are located in the vicinity of the body walls), we retain the $B_m=\SI{3}{\meter}$ CFD mesh for future computations (\ie meshing a length of $4.5A_m$ on each side of the body). %

\subsubsection{Sensitivity to tolerance targets of the FD method }
\label{sec:sensitivityResultTolerance}
A study was conducted on the tolerance targets and their effects on the obtained results within the FD approach. It was shown that using the original tolerance ($10^{-9}$, $10^{-8}$ and $10^{-7}$ for the velocity, pressure and \ofkw{PIMPLE} loop, respectively) leads to a large computational cost, mainly because they correspond to dimensionless quantities, obtained after normalization by representative field values. 
In the FD approach, the field of interest, especially the pressure $p^*$, is order of magnitude smaller that the resolved fields in the DD approach. In practice it was found that the targets could be multiplied by $10^4$ (\ie $10^{-5}$, $10^{-4}$ and $10^{-3}$ for the velocity, pressure and the \ofkw{PIMPLE} loop, respectively), with a maximal difference on the obtained complementary pressure field after the simulation of 6 wave periods of only $3.6\%$. 
We would like to underline that this difference is relative to the \emph{complementary} pressure field amplitude, which is an order of magnitude lower than the total dynamic pressure field ($\approx12\si{Pa}$ and $700\si{Pa}$ respectively, see \eg \cref{fig:sub:plotOverLineMultiCasevCvsdCP}).

\section{Validation against experimental measurements}
\label{sec:resultsVsLiterature}
\subsection{Introduction}
In this section, we apply the developed coupling schemes to a range of incident regular wave conditions for the submerged horizontal cylinder described in \cref{sec:casedescription} and \cref{sch:cylinderVenugSch3}, keeping the same geometry: $H_c/L_c=1/2$, $d_c/H_c=4.1$ and  $h/d_c=2.68$. This set-up was considered experimentally by \textcite{arai1995forces,venugopal_hydrodynamic_2002,venugopal_drag_2009}, and simulated numerically by \textcite{li_hydrodynamic_2010} using a CFD VoF approach.



While in the current work the wave period remains fixed at $T=\SI{2}{s}$, several increasing wave heights were simulated leading to a wave steepness varying from $H/\lambda =0.5\%$ to $8.6\%$. The Keulegan-Carpenter number, defined as
\label{sec:KCdefinition}
\begin{equation}
    \text{KC}=           \pi \dfrac{H }{L_c} \dfrac{ \cosh(k(h-d_c))}{\sinh(kh)}, 
    \label{eq:KC}
\end{equation}
where $k=2\pi/\lambda$ is the wave number, represents the ratio of the amplitude of horizontal excursion of a fluid particle by the horizontal length of the body, assuming linear waves. %
KC will thus be used to parameterize the wave height, and varies as $\text{KC}\in[0.11, 2.07]$. %
Note that both wave steepness and KC vary by a factor of $\sim 18$ from the lowest wave height to the most nonlinear condition.

To get a more synthetic appraisal of models' results and compare them with the above mentioned reference data sets, we restrict our attention to the values of the hydrodynamic coefficients (HC) extracted from the time series of computed loads on the body as explained in the next sub-section.

\subsection{Computation of the hydrodynamic coefficients (HC)}
Literature comparisons are usually done on the hydrodynamic coefficients (namely the inertia coefficient $C_{Mx,z}$ and drag coefficient $C_{Dx,z}$) that model the forces applied on the body following the so-called Morison equation (\textcite{morison_force_1950}):
\begin{subequations}
\begin{align}
    F_{xm} & = \frac{1}{2} \rho C_{Dx} H_c u_x \sqrt{u_x^2 + u_z^2}   + \rho A C_{Mx} \dot{u}_{x} \label{eq:HydrodynamicCoefficients:fx}\\
    F_{zm} & = \frac{1}{2} \rho C_{Dz} L_c u_z \sqrt{u_x^2 + u_z^2}   + \rho A C_{Mz} \dot{u}_{z} \label{eq:HydrodynamicCoefficients:fz}
\end{align}
\label{eq:HydrodynamicCoefficients}
\end{subequations}
where $F_{xm}$ and $F_{zm}$ are the modeled (by the Morison formula) horizontal and vertical loads applied on the body (expressed in $\si{\newton\per\meter}$ in the present 2D framework), and $A=H_cL_c$ is the area of the rectangular cylinder cross-section. %
For simplicity, those expressions of forces components will be called the Morison loads hereafter. 
$u_x$, $u_z$ are respectively the horizontal and vertical velocity of the inflow, and $\dot{u}_{x}$, $\dot{u}_{z}$ are the accelerations of the inflow in the corresponding directions.  Note that the selection of those kinematic values represents a first significant decision.
The Morison equations were originally developed to model the force applied by a uniform oscillatory flow on an object of small dimensions (relative to the wavelength). The associated assumptions is that the body is subjected to an inflow on which it has a limited impact. In this case, the choice of the associated kinematic is straightforward: one should select the kinematics of the unperturbed flow (\ie in the absence of the body). 
By extension, the imposed wave model at the inlet on the computational grid (here a stream function theory) is applied at the location of the center of the cylinder. %
Given a temporal load series, many methods exist to estimate the ``corresponding'' hydrodynamic coefficients, \ie the hydrodynamic coefficients that yield Morison loads as close as possible to the measured or simulated loads. %
One of the most common methods is to minimize the square root of the error between the modeled loads and the real ones. At a given time step $i$ the error is given by $e_{x,z}(i) = F_{x,z}(i) - F_{xm,zm}(i) $. 
Indexes $x,z$ represent the errors for the horizontal and vertical components of the force, respectively. Thus, one could minimize the resulting mean (over time) square error:
\begin{equation}
     L_2(F_{x,z})=\frac{\sqrt{\frac{1}{N}\sum_i^N e_{x,z}^2(i)}}{\max_i  F_{x,z}(i)  - \min_i  F_{x,z}(i) }. 
    \label{eq:L2_error}
\end{equation}

Following \textcite{venugopal_hydrodynamic_2002}, the coefficients that minimize these $L_2$ errors over a certain number of time steps $N$ can be analytically derived as:
\begin{subequations}
\begin{align}
    C_{Dx,z} & = \dfrac{2}{\rho D} \dfrac{f_1f_2 -f_3f_4}{f_2f_5- f_4^2} \label{eq:HydrodynamicCoefficientsWithfi:fx}\\
    C_{Mx,z} & = \dfrac{2}{\rho A}\dfrac{f_3f_5 -f_1f_4}{f_2f_5- f_4^2}  \label{eq:HydrodynamicCoefficientsWithfi:fz}
\end{align}
\label{eq:HydrodynamicCoefficientsWithfi}
\end{subequations}
where $D$ is either $H_c$ or $L_c$ for the horizontal and vertical drag coefficients respectively, and the $f_i$ terms are given by:
\begin{equation}
\begin{aligned}
    f_1 & = \sum_i^N F_{x,z}(i) u_{x,z}(i) \norm{\tI{u}(i)} \\
    f_2 & = \sum_i^N \dot{u}_{x,z}^2(i) \\
    f_3 & = \sum_i^N F_{x,z}(i) \dot{u}_{x,z}(i)  \\
    f_4 & = \sum_i^N u_{x,z}(i) \norm{\tI{u}(i)} \dot{u}_{x,z}(i)  \\
    f_5 & = \sum_i^N u_{x,z}^4(i)  \\
\end{aligned}
\label{eq:fiforhydrocoeffs}
\end{equation}
\begin{rmk}
    Note that another method is presented in \textcite{arai1995forces}: it consists in expanding in terms of Fourier series the kinematic variables - based on their analytic expressions - to obtain the Fourier decomposition of the Morison loads. Then, a numerical harmonic decomposition is also performed on the obtained loads. Finally, both decompositions are equalized, and the corresponding terms are identified. 
    With this method, the coefficients are obtained in terms of the Fourier harmonics that constitute the load series. This method was also implemented in this work, with no major differences in terms of obtained HC. The results will thus not be presented here.
\end{rmk}

\subsection{Results and discussion}
\begin{figure}[htbp!]
        \centering
        \includegraphics{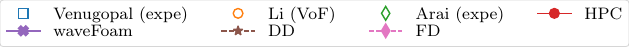}
        \begin{subfigure}[c]{0.5\textwidth}
                    \includegraphics{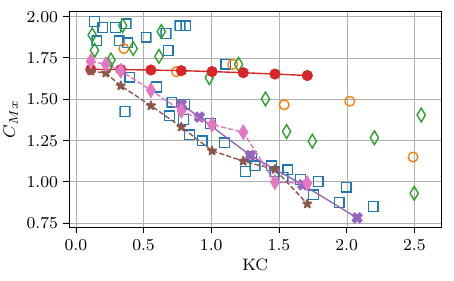}
                \caption{}
                \label{fig:sub:plotCdxCdxVenugopalArticlevCdCCmx}
        \end{subfigure}%
        \begin{subfigure}[c]{0.5\textwidth}
                \includegraphics{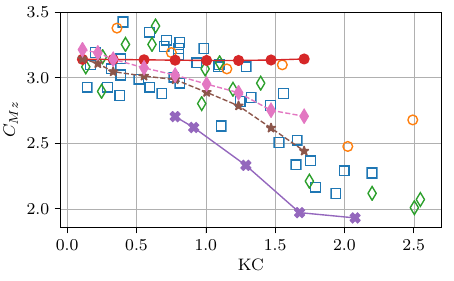}
                \caption{}
                \label{fig:sub:plotCdxCdxVenugopalArticlevCdCCmz}
        \end{subfigure}%

        \begin{subfigure}[c]{0.5\textwidth}
                \includegraphics{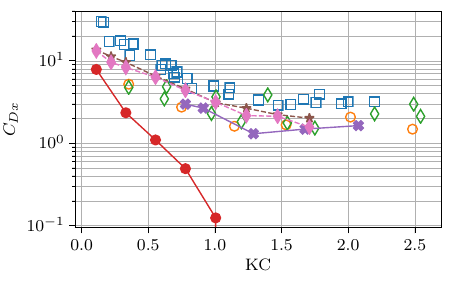}
                \caption{}
                \label{fig:sub:plotCdxCdxVenugopalArticlevCdCCdx}
        \end{subfigure}%
        \begin{subfigure}[c]{0.5\textwidth}
                \includegraphics{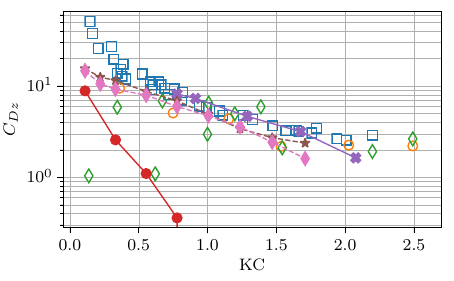}
                \caption{}
                \label{fig:sub:plotCdxCdxVenugopalArticlevCdCCdz}
        \end{subfigure}%

        \begin{subfigure}[c]{0.5\textwidth}
                \includegraphics{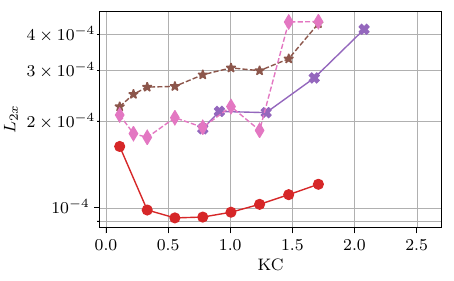}
                \caption{}
                \label{fig:sub:plotCdxCdxVenugopalArticlevCdCErrx}
        \end{subfigure}%
        \begin{subfigure}[c]{0.5\textwidth}
                \includegraphics{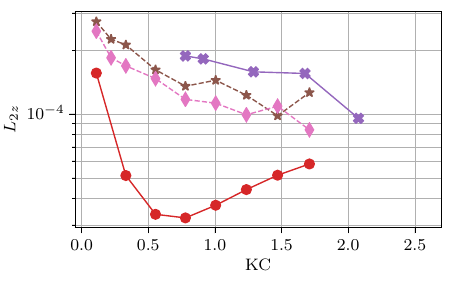}
                \caption{}
                \label{fig:sub:plotCdxCdxVenugopalArticlevCdCErrz}
        \end{subfigure}%

        \caption{Inertia and drag coefficients in the horizontal ($x$) and vertical ($z$) directions obtained with the DD coupling method, the FD coupling method, the VoF-FVM method (waveFoam) and the HPC method for different KC numbers. The $L_2$ error of the Morison fitting to the temporal loads series is also shown in the lower panels. Large errors reveal that the Morison model is not well adapted to describe the temporal series of loads.}
        \label{fig:plotCdxCdxVenugopalArticlevCdC}
\end{figure}

All results in terms of HC are depicted on \cref{fig:plotCdxCdxVenugopalArticlevCdC}:  \cref{fig:sub:plotCdxCdxVenugopalArticlevCdCCmx,fig:sub:plotCdxCdxVenugopalArticlevCdCCmz} show respectively the obtained horizontal and vertical inertia coefficients ($C_{Mx}$ and $C_{Mz}$ respectively), \cref{fig:sub:plotCdxCdxVenugopalArticlevCdCCdx,fig:sub:plotCdxCdxVenugopalArticlevCdCCdz} the obtained drag coefficients ($C_{Dx}$ and $C_{Dz}$ respectively), and \cref{fig:sub:plotCdxCdxVenugopalArticlevCdCErrx,fig:sub:plotCdxCdxVenugopalArticlevCdCErrz} the $L_2$ norm of the error (given by \cref{eq:L2_error}) in the reconstruction of the obtained loads \emph{via} the Morison formula \cref{eq:HydrodynamicCoefficients:fx,eq:HydrodynamicCoefficients:fz}. We would like to stress that this error does not represent a discrepancy between the obtained values and the experimental ones, but only the consistency in representing the loads with the Morison.  Note that, in order to make the comparison easier, the axis ranges of all panels of this figure are chosen so as to fit the range of the presented data. 

\paragraph{\textbf{HC from HPC standalone simulations}}~~\\ 
The HC that model the potential loads (from FNPF-HPC model), represented on \cref{fig:plotCdxCdxVenugopalArticlevCdC}, are consistent with the potential assumptions: 1) drag coefficients are of very small amplitudes when compared to experimental ones, 2) at small KC, \ie when the viscous and vorticity effects are expected to be of lower magnitude, the inertia coefficients exhibit good agreement with the experimental ones, and 3) the magnitude of variation of the inertia coefficients with KC is relatively small, denoting that the first harmonic of the load (linear with respect to the wave height, and thus with KC) remains dominant.

\paragraph{\textbf{HC from \ofkw{waveFoam} standalone simulations}}~~\\ 
The HC obtained with the FVM-VoF method alone yield a good agreement with the experimental results, except for the vertical inertia coefficient (\cref{fig:sub:plotCdxCdxVenugopalArticlevCdCCmz}). This discrepancy is the direct consequence of the observed difference in terms of vertical load amplitude shown in \cref{sec:temporalloads}, \cref{fig:plottemporalLoadsHPCOFvCdCwF} and is thought to be caused by an artificial numerical damping of the incoming wave  during the propagation phase. Note that because a phase shift of the wave kinematics was also observed during the propagation, an \emph{ad hoc} correction was applied before computing the HC. 

This further grants confidence in the fact that the coupled approaches, as well as HPC, correctly capture the vertical loads despite exhibiting some differences with the \ofkw{waveFoam} simulation. 

\paragraph{\textbf{HC from the DD and FD coupled simulations}}~~\\ 
Despite the fact that the BC are enforced from the FNPF-HPC results, the drag force is correctly recovered with both the DD and FD approaches for the whole studied range of KC. In the same manner, the evolution of the inertia coefficients is recovered: at small KC values, the potential (HPC) inertia coefficients are recovered, while for larger values the curves separate and a good agreement is maintained with the inertia coefficients from \textcite{venugopal_drag_2009}.

\subsection{Computational efficiency} 
In this section, few CPU costs are compared on the main studied case ($T=2$~s, $H/\lambda = 3.5\%$). Almost all computations were run on a 40 core Intel\textregistered\ Xeon\textregistered\ CPU E5-2650 v3 @ 2.30GHz. We took advantage of the parallelism only in waveFoam standalone computations. 

\Cref{tab:CPUcosts} shows some of the lower achieved CPU times needed to obtain the presented results with the 3 RANS methods (waveFoam, DD, FD). The mesh is the same in terms of discretization ($dx=\SI{0.056}{\meter}$ in the body vicinity) but its span, particularly in the stream-wise direction, is different, respectively \SI{24}{\meter} for waveFoam (\ie the full extent of the NWT) and $B_m=$ \SI{3}{\meter} for the coupled methods. In \cref{sec:mesh_breatdh}, it was shown that a reduction of this CFD extent below \SI{3}{\meter} is possible with the DD method, %
but, as no tests regarding the mesh extent $B_m$ were conducted with the FD method, CPU times using this larger extent are shown instead, for a fair comparison between models.  

\begin{table}[htb!]
    \newcommand{\mrf}[1]{\multirow{1}{*}{#1}}
    
    \centering
    \scriptsize{
    \begin{tabular}{|l|c|c|c|c|}
                                             & waveFoam     & DD coupling & FD coupling  & HPC \\
 periodic behavior reached after $\dfrac{t-t_0}{T}=$ & 15           & 6              & 6  & 15\\
 number of cells                              & \SI{138e3}{} & $<$\SI{31e3}{} & $<$\SI{31e3}{} & \SI{14e3}{}\\
 CPU time to reach periodic behavior                   & $\sim$\ti{24}{04}      & \ti{3}{11}          & \ti{3}{00} & $\sim$\ti{1}{25} \\
 CPU time for one additional wave period                   & \ti{1}{46}        & \ti{0}{36}          & \ti{0}{30} & $\sim$\ti{0}{05}\\
     \end{tabular}
   }
   \caption{CPU costs (the VoF models use the same spatial discretization in the body vicinity, $dx=\SI{0.056}{\meter}$) using one computational core. The CPU values for the coupled approaches do not include the CPU time of the HPC simulation, which is given in the last column. Note that those CPU time values might not be perfectly accurate (see text for details).
   }
     \label{tab:CPUcosts}
 \end{table}

 Moreover, the presented CPU values for waveFoam are estimated (see ``$\sim$'') by comparing a parallel run on 4 cores and a sequential one over 1 wave period, and thus might not be perfectly accurate. Moreover, these values also proved to be prone to differences even when the same computation is performed (for example the frequency of writing results on files can be a source of variation) and trust can only be granted in their orders of magnitude, say $\pm10\%$. Lastly, no investigation of the requested tolerance was conducted for \ofkw{waveFoam} or the DD method, and even the one concerning FD approach could have been pushed further (see \cref{sec:sensitivityResultTolerance}).

 Nevertheless, \cref{tab:CPUcosts} shows that a significant gain is achieved with the coupling approaches: the computational time per wave period is divided by $\sim 3$ when compared with \ofkw{waveFoam}. If one is interested in the periodic behavior, the required CPU time is divided by $\sim 8$. Valuable improvements of the coupling methods and the potential model can thus be obtained for a relatively contained CPU cost. 
 Moreover, performing mesh/time step sensitivity studies with these coupled methods would be highly beneficial due to the separation of wave propagation and body vicinity physics, but also the reduced CPU cost.

\section{Conclusions and outlook} \label{sec:conclusion}
Two coupling strategies have been developed and validated within the \openfoam framework, namely a domain decomposition (DD) and a functional decomposition (FD) approach, using a FNPF model as a driver for the large scale flow. One specificity of the present work is that the potential flow solution, obtained with an accurate and efficient Harmonic Polynomial Cell (HPC) method, is computed in the presence of the body. Both coupling strategies are currently one-way schemes, implying information is only exchanged from the potential model to the CFD one. While the unidirectional property of the coupling restricts the usage to cases where vorticity, viscous and turbulent effects are restricted to an area of limited extent around the body, purely potential diffraction effects are taken into account at the early stage of the solution with the potential flow solver. 

The approaches were applied on a 2D wave-structure interaction case which exhibits common features with ocean engineering applications, in particular the presence of sharp corners. The above mentioned assumptions were shown to be fulfilled on this wave body interaction case: a rectangular-shaped fixed cylinder immersed below the MSL in regular nonlinear waves.
 We have shown that not only loads exerted on the body, but also local field descriptions such as flow vorticity, turbulent viscosity, etc. are correctly captured by both coupling methods as complementary results using as input potential flow model predictions, and closely match the results obtained \emph{via} an independent standalone CFD simulation.
Furthermore, the coupled simulations showed a strong reduction of the CPU burden. For example, the flow vorticity field in turbulent conditions is well recovered in the body vicinity, though the total domain span is divided by more than 8 when compared to the requirements of a standalone \openfoam computation.

Afterwards the effect of the variation of the incident wave height was studied, by covering a range of values of the KC number, and the obtained hydrodynamic coefficients (HC) of the considered body, used to model the obtained loads in a synthetic manner, were computed and compared with HC from experimental time series of measured loads. 

Even though the horizontal and vertical inertia coefficients can be captured in a relatively accurate manner by the potential model for small KC, the drag coefficients cannot be recovered with such method. 
In the same manner, larger KC numbers trigger effects laying outside of the potential assumptions, significantly affecting the load values. The coupled approaches, making use of the potential wave propagation, yield a significant improvement: all inertia and drag HC are correctly estimated for the complete range of KC values considered here. Moreover, the potential model is shown to accurately capture the wave propagation phase, while setting up a FVM-VoF approach able to simulate accurately both the propagation but also the small scale effects near body's walls proved to be challenging. Finally, a great improvement in terms of CPU cost is achieved with both the coupling schemes. 

Note that these conclusions are drawn on a particular configuration, for which the assumption of the one-way coupling seems to hold: the retroactive effects of the viscous and turbulent effects on the large scale flow are limited. Let us recall that, on the contrary, no assumption of the smallness of purely potential diffraction effects needs to be done, as the potential solution is simulated taking into account the presence of the body.

Overall, it is thought that using such methods applied at a local scale around the body (or each individual body in a multi-body case) as a further step after a larger scale potential flow computation are of great interest. In addition to the significant reduction of the total CPU time, some of the other benefits are: (i) a first estimation of loads of the body is already available after the potential flow simulation, (ii) these approaches allow to identify and quantify the non-potential effects on both the flow field and loads on the body, (iii) the quality and accuracy of the propagated wave field is high, without effect of artificial numerical damping, which may become significant for CFD simulations over long distances, (iv) all steps of work dealing with the calibration of the CFD model and sensitivity studies of various parameters are much easier and faster as the CFD computational domain is drastically reduced, (v) a larger number of wave conditions can be simulated or explored for a given computational burden.

Several continuations of this work can be drawn, most of them being common between the DD and FD coupling approaches.
The FNPF-HPC model is currently developed in 2D. However, respecting the \openfoam philosophy, none of the presented implementations use a 2D hypothesis. Thus, provided a 3D external model, the extension of the current coupling methods to 3D cases should be straightforward.
While not shown here, first tests of the coupling methods have been conducted with a free surface piercing body. However, due to the one-way nature of the couplings, the current implementations proved to be prone to instabilities in some cases. More complicated matching strategies could be setup to overcome this difficulty. For example, \textcite{zhang_multi-model_2018} applied relaxation techniques close to the outer boundaries of the viscous domain to allow for a smoother enforcement.
More generally, this issue could certainly be better resolved by extending to a two-way coupling. Within a two-way coupling, the turbulent flow also acts on the computation of the potential flow itself. This method would also ease the matching with the potential free surface, probably removing the need of invoking any smoothing. 

\section*{Declaration of competing interest}
The authors declare that they have no known competing financial interests or personal relationships that could have appeared to influence the work reported in this paper.
\section*{Acknowledgment}
This work was partly supported by the École Normale Supérieure de Cachan (ENS Cachan, France) in the form of a Ph.D. grant attributed to Fabien Robaux. This research did not receive any other specific grant from funding agencies in the public, commercial, or not-for-profit sectors.

\appendix


\printbibliography

\end{document}